\documentclass[a4paper,11pt]{article}
\pdfoutput=1 
\usepackage{jheppub} 
\usepackage{amsmath,amssymb,mathtools,tabu}
\usepackage{epsfig,epstopdf}  
\usepackage{graphicx}
\usepackage{slashed}             
\usepackage{url}
\usepackage{color,xcolor}
\usepackage{multirow}
\usepackage{placeins}
\usepackage{xspace}
\usepackage{appendix}
\allowdisplaybreaks

\bibpunct{[}{]}{,}{n}{}{,}

\preprint{TIFR/TH/19-33}
\abstract{We propose a method to identify jets consisting of all the visible remnants of
a boosted top quark decaying semileptonically with an electron in the final state.
An overlap of electron shower with the $b$ quark initiated shower, and the large
nontrivial energy-momentum carried by the invisible neutrino in the top quark decay are the two
obstacles to achieving this aim. Our method uses the distribution of energy in different 
parts of the detector to identify a jet containing an energetic electron, involves use of
substructure of the jet to determine the momentum associated with the electron and then completes
the identification of top jet with the construction of new variables which would reflect the top 
quark decay kinematics. The last part involves an ansatz of the existence of a massless, invisible 
four-momentum \emph{roughly} collimated with the electron, whose four-momentum when combined with 
that of the  the electron and the full jet, reconstructs to the $W$ boson and the top quark respectively. We demonstrate the efficacy of this proposal  using simulated data and show that our method not only reduces the backgrounds from light flavor jets, $b$ jets from QCD, and hadronic top jets, it can also tell apart jets rich in electrons but not due to top quark decays.
}

\begin{document}
\title{Jets with electrons from boosted top quarks}

\author[a]{Suman Chatterjee,}
\emailAdd{suman.chatterjee@tifr.res.in}
\author[b]{Rohini Godbole,}
\emailAdd{rohini@iisc.ac.in}

\author[c]{and Tuhin S. Roy}
\emailAdd{tuhin@theory.tifr.res.in}

\affiliation[a]{Department of High Energy Physics, Tata Institute of Fundamental Research, \\1 Homi Bhabha Road, Mumbai, 400005, India}
\affiliation[b]{Center for High Energy Physics, Indian Institute of Science, Bangalore 560012, India}
\affiliation[c]{Department of Theoretical Physics, Tata Institute of Fundamental Research,\\1, Homi Bhabha Road, Mumbai, 400005, India}

\maketitle

\section{Introduction}\label{sec:intro}
Identifying events containing remnants of top particles~\cite{Abe:1995hr}  produced   at the  Large Hadron Collider (LHC) is of paramount interest, since they provide  profound insights into the standard model (SM) of particle physics as well as allow us to look for the hints of new physics. In fact, models which address the naturalness problem in the SM, or attempt to solve the issue of flavor hierarchies in SM fermions, often contain new heavy states which couple to top quark. Consequently, traces of new physics, if it exists, are very likely to show up  in events with top quarks.  Note,  however, that top physics at LHC is significantly different from that at older colliders like Tevatron.   One of the features of the high center-of-mass energy associated with proton-proton collisions at the LHC is that, copious amounts of top quarks are produced with transverse momenta $p_T \gg m_t$, where $m_t$ denotes the top quark mass, even if the production processes are entirely due to SM. In case of top particles resulting from decays of heavy states of new physics, high $p_T$ is guaranteed.  Because of high boost factors, these  top quarks decay into collimated collections of particles that often look like single jets of large size in the detector. Tagging  these top jets, which contains all the decay products of hadronically decaying top quarks is quite a mature field.  A plethora of tagging algorithms  have been proposed which range from the substructure analyses~\cite{Thaler:2008juleptop,Kaplan:2008ieJHU,Abdesselam:2010pt,Plehn:2011sjHEPTTv1,Larkoski:2013eya,Altheimer:2013yza,Ellis:2012sn,Soper:2012pb,Ellis:2014eya,Kasieczka:2015jmaHEPTTv2,CMS-PAS-JME-15-002,Aaboud:2018psm} to methods taking full advantage of recent advances in the machine learning~\cite{Cogan:2014oua,Almeida:2015jua,deOliveira:2015xxd,Baldi:2016fql,Butter:2017cot,Komiske:2017aww,Louppe:2017ipp,Kasieczka:2017nvn,Komiske:2018cqr,Collins:2018epr,CMS-PAS-JME-18-002,Collins:2019jip,Qu:2019gqs,Roy:2019jae}.

Tagging jets containing decay products of  semileptonically decaying top particles, and identifying the momentum of the lepton from top decay is probably even more interesting in terms of extracting physics associated with the top quark. Note that since the top quark decays before it can hadronize, the energy and angular distributions of its decay products carry imprints of the spin direction of the decaying top quark -- the correlation of these distributions with the top polarization being decided by the chiral structure of the top quark decay interactions (see for example,~\cite{Jezabek:1994qs,Bernreuther:2008ju}).
In case of semileptonic decays, the angular distribution of the lepton (inside the jet) is an unambiguous probe for the top quark polarization as it is independent (to the leading order) of the anomalous part of the top decay vertex~\cite{Grzadkowski:1999iq,Rindani:2000jg,Grzadkowski:2001tq,Grzadkowski:2002gt,Godbole:2006tq,Godbole:2010kr}.   
 Top quark polarization is determined by the chiral structure of the  interaction responsible for its production.  Indeed the  polarization of the top jet produced via the SM processes is precisely predicted : zero for the dominant $t \bar t$ production via the vector QCD interaction and a nonzero, but precisely predicted value for the  EW, single top production. Hence any deviation away  from these SM predictions  of the top jet polarization, measured using the angular distributions of the decay lepton  will be a telltale signature of a beyond SM (BSM) mechanism giving rise to non-SM values of  top polarization (see for example ~\cite{Godbole:2010kr,Belanger:2012tm,V.:2016wba}).
Furthermore, the energy distribution of the lepton from polarized top quark carries nontrivial information about the anomalous top decay vertices, see for example~\cite{AguilarSaavedra:2010nx, Prasath:2014mfa, Jueid:2018wnj}, and consequently,  provides a means to detect physics beyond SM.  
Further, spin-spin correlations for top quarks produced in pairs or in association with a Higgs boson, also carry information about the chiral structure of the interactions responsible for the production which are  faithfully reflected in the angular correlations between the decay leptons. 
The need for a dedicated method that identifies these semileptonic top quark jets and, more importantly, efficiently gives the energy and the orientation of the lepton inside the jet can not be overstated.

Unfortunately, neutrinos associated with these semileptonic decays  carry away energy and mass of top. This  makes tagging these semileptonic top jets  highly nontrivial.   Not surprisingly, tagging top jets where boosted top quarks decay to leptons has not received that much attention.  In case of top quark decays to muons,  these  ``muonic top jets" get characterized in terms of non-isolated muons which leave tracks in the muon spectrometer portion of the detector. Matching the spectrometer track to a track seen at the tracker portion of the detector may allow us to reduce the background due to QCD heavy flavor jets.  Indeed, some of the early proposals such as Ref.~\cite{Thaler:2008juleptop,Plehn:2011tf}  use the muon track inside the jet. Other relevant proposals which can be useful in this case involve the so-called``mini-isolation" associated with the muon~\cite{Rehermann:2010vq}, or the lepton energy fraction in a smaller subjet~\cite{Brust:2014gia,Agashe:2018leo} within a large sized jet.  Note, however, that tagging ``electronic top jets" where the boosted top quark decays to an electron is rather hard, since identifying an electron when its shower partially overlaps with the shower from the fragmentation of a $b$ quark is difficult. Special reconstruction of electron tracks, for example using Gaussian-sum-filter (GSF)~\cite{Adam:2003kg} developed by CMS~\cite{CMS_ex}, enhances the efficiency of electron reconstruction significantly, but it comes only at the cost of high rates at which  pions fake electrons, especially, at high momenta~\cite{Sirunyan:2017ulk}. Another difficulty associated with  tagging semileptonic top jets is the fact that event level information such as total missing energy of the  event can not be used. This seems to suggest that very little kinematic information associated with top quark decays can be  of use.

In this work we propose a dedicated tagger for electronic top jets. The procedure of identification relies on two different kinds of observables and an algorithm. The first set of these variables is a collection of jet substructure based observables which  attempt to decide whether the observed jet is consistent with a jet containing an energetic electron or not, by drawing information from various parts of the calorimeters and the tracker.  The second set of observables are calculated using an ansatz that there exists a non-zero momentum four-vector (massless and approximately collimated to the candidate  $4$-momentum for the electron) which reconstructs $W$ boson mass when combined with the electron,  and reconstructs top quark mass when combined with the full jet. These observables therefore have physical interpretations only when the jet is an electronic top jet. Clearly, the efficacy of the second set relies on finding the candidate  $4$-momentum for the electron from top quark decay. The purpose of our proposed algorithm is to find a candidate for the electron given a jet by combining the tracker and the calorimeter information.  

We benchmark the performance of our proposal by using simulated events for top jets with $p_{T}>500~\text{GeV}$. In particular, we  manage to identify electronic top jets at quite a high efficiency while at the same time suppressing backgrounds from heavy and light flavor QCD jets, as well as jets from hadronically decaying top quarks.  One of the novel aspects of our proposal is that since we utilize more kinetic information associated with top quark decays, we are able to tell apart the electronic top jets from those jets which contain showers due to a $b$ quark and an energetic electron  but are not due to top quark decays. As a proof of principle, we consider a new physics example. To be specific, we take jets containing decay products of the stop quark of supersymmetry~\cite{Martin:1997ns}, where the stop quark decays to a $b$ quark, an electron, a neutrino, and an invisible neutralino. Even though we do not use any information pertaining to these stop jets for constructing the tagger, we show that our method can separate  electronic top jets from stop jets quite efficiently.  Based on these observations we propose a simple extension of the tagger, which apart from finding electronic top jets, can also identify jets due to new physics processes as anomalous jets. 

Even though we concentrate our efforts in identifying  electronic top jets, our proposal, after slight modifications, can at the same time tag muonic top jets as efficiently as electronic top jets.  
By construction, the distribution of the variables we propose in this work are identical whenever the resultant top jets are rich in muons or in electrons.
The only change needed is in the procedure of identifying the muon candidate inside the jet, which  we outline towards the end of this work in the Appendix~\ref{sec:appendix1}.  

The rest of this paper is organized as follows: in Sec.~\ref{sec:method} we describe the methodology, in  Sec.~\ref{sec:sample} we  give details of the simulation we performed in order to benchmark our proposals, in Sec.~\ref{sec:results} we show the results,  and finally in Sec.~\ref{sec:conclusion} we conclude.

\section{Method}\label{sec:method}
The purpose of this section is to chalk out a strategy that can identify whether a jet is consistent with a boosted top quark where the top quark decays to an electron.
Broadly speaking, the central part of this identification is broken down into following steps.

\begin{enumerate}	
\item The first stage of this procedure is to groom the jet using ``soft drop'' (SD) method~\cite{Larkoski:2014wba}, which allows us to  
identify the last stage of clustering with hard splitting, and to remove soft radiation making it somewhat robust under underlying event and pileup~\cite{CMS-PAS-JME-14-001}. 

\item The distribution of energy of the groomed jet, as well as of its subjets, over different parts of the calorimeters (namely, the electromagnetic portion or the ECal and the hadronic portion or the HCal) and the signature left at the tracker allow us to determine whether the jet contains an energetic electron within itself or not. We use the notation $\mathcal{V}_e$ to denote the set of all variables, which we employ to determine whether the jet under consideration is consistent with a jet containing a hard electron. 

\item We devise a set of steps that starts with the constituents of the groomed jet and determines the momentum four-vectors, which may correspond to the momenta of the $b$ quark and the electron from top quark decay.
 
\item The ansatz that the jet contains the remnants  of a boosted top implies that if a  momentum four-vector  (representing the invisible neutrino), \emph{massless  and roughly collimated} to the electron, is added to the electron or  the full  jet, the resultant should reconstruct the momentum of a $W$ or a  top particle respectively.  As we show later in this section, this ansatz allows us to determine the energy carried away by the neutrino and some aspects of its direction inside the jet; we develop few measurable variables using this ansatz, and refer this set of variables based on neutrino properties by $\mathcal{V}_\nu$.

\item We construct two boosted decision trees (BDT)~\cite{5392560}, denoted by $\mathcal{B}_{e}^{t/b}$ and $\mathcal{B}_{\nu}^{t/b}$ which use the variables $\mathcal{V}_e$ and $\mathcal{V}_\nu$ respectively as input. 
Both these BDTs are optimized to separate a sample of jets containing the decay products of top quarks from a sample of jets initiated by $b$ quarks. 
After the training, any jet gets characterized by two BDT responses.
In other words, the jet gets mapped to a point in a plane of BDT responses. 
We identify and thereby \emph{veto} regions in this plane which are dense in $b$ quark initiated jets. As we show later, this procedure yields an efficient tagger for boosted tops decaying to electrons.
\end{enumerate}

In the rest of this section we give brief account of all the five steps mentioned before. 

\subsection{Grooming }

In the first stage, we groom the jet under consideration. Typically, input jets result from infrared- and collinear-safe clustering algorithms, such as $k_T$~\cite{Ellis:1993tq}, Cambridge-Aachen (C$/$A)~\cite{Dokshitzer:1997in}, anti-$k_T$~\cite{Cacciari:2008gp} etc with a given distance parameter $R$.
In this work, we use soft drop to groom the original jet $J$. 
Here we provide a short description of the algorithm following Ref.~\cite{Larkoski:2014wba}.
The first step towards this is to recluster the constituents of the jet using C$/$A algorithm and a very large distance parameter ($\gg R$) which returns the same jet but gives a C$/$A clustering history for the jet.  The algorithm to find the groomed jet is given below.
\begin{enumerate}
	\item Undo the last step of the clustering of $J$, which  splits it into two subjets (say $j_1$ and $j_2$, with  $j_1$ being the more energetic than $j_2$). 
	\item If $j_2$ is sufficiently hard, grooming  ends. The hardness criterion is given as  
	\begin{equation}
	p_{T j_2} \  \geq \  z_{\text{cut}} \left(\frac{\Delta R_{12}}{R} \right)^{\beta} \times (p_{T j_1}+p_{T j_2})  \; ,
	\label{Eq_SD}
	\end{equation} 
	
	where $p_{T j_1}$ and $p_{T j_2}$ are $p_{T}$ of subjets $j_1$ and $j_2$ respectively, $\Delta R_{12}$ is the distance between $j_1$ and $j_2$ in rapidity-azimuth plane, and finally $z_{\text{cut}}$ and $\beta$ are taken to be $0.1$ and $0$ respectively. 
	
	\item 	If $j_2$ is too soft w.r.t. $J$ such that it fails the criterion in Eq.~\eqref{Eq_SD}, the subjet $j_2$ is discarded and  the whole procedure is repeated after replacing $J$ by $j_1$.   
\end{enumerate}

After grooming ends, we identify the remnant jet $J$  to be the groomed jet. For all practical purposes we discard the original jet for the rest of this work, and use only the groomed jet.  Apart from the four-vector representing the groomed jet, it is also characterized by the subjets $j_1$ and $j_2$.

\subsection{Variables for tagging}
\label{Sec_Ve}

In this subsection, we provide definitions of the variables in the set $\mathcal{V}_e$. The exact methodology via which we estimate these variables will be provided later in Sec.~\ref{Sec_Results_Ve}.

\begin{itemize}
\item  The invariant mass of the jet $J$ turns out to be one of the most useful variable. In this work we use the notation $m_{\text{SD}}$ to denote its mass which reflects the fact that $J$ is already groomed via soft drop algorithm. 

\item We denote the hadronic energy fraction and the electromagnetic energy fraction of a jet $J$ by $f_\text{h}$ and $f_\text{em}$ respectively.  Apart from measuring these quantities for the full jet, we also consider these energy fractions for subjets of $J$ also.   
We set up notations $f_\text{h}^j$ and $f_\text{em}^j$, to be used later on, to denote  the hadronic and electromagnetic fractions of the total energy of the $j$-th subjet respectively.
All these fractions are  thus given by:
\begin{equation}
\begin{split}
f_\text{h} \ \equiv \ \frac{1} {E_J} \: \sum_{k \in \text{HCal}} E_J^{(k)} \; \quad \text{and} \quad
f_\text{h}^j \ \equiv \ \frac{1} {E_j} \: \sum_{k \in \text{HCal}} E_j^{(k)}	\; , \\
f_\text{em} \ \equiv \ \frac{1} {E_J} \: \sum_{k \in \text{ECal}} E_J^{(k)} \; \quad \text{and} \quad
f_\text{em}^j \ \equiv \ \frac{1} {E_j} \: \sum_{k \in \text{ECal}} E_j^{(k)}	\; ,
\label{hcaleq}
\end{split}
\end{equation}  
where $E_J$ ($E_j$) and  $E_J^{(k)}$ ($E_j^{(k)}$) represent the total energy and the energy of the $k$-th constituent of the jet $J$ (subjet $j$)  respectively.
We also use nonhadronic energy fraction, labelled by $f_\text{1-h}$, and defined as
\begin{equation}
 f_\text{1-h} \ \equiv \ 1-f_\text{h} \; .
\label{Eq_fnh}
\end{equation}  
The variable $f_\text{1-h}$ is same as $f_\text{em}$ if the jet does not have any muon as constituent, otherwise they differ by the energy fraction carried by the muons inside the jet.

\item  We propose a new variable which measures the asymmetry between hadronic energy deposits of two SD subjets $j_1$ and $j_2$. 
\begin{equation}
{A}_h \ \equiv \ \frac{(f_\text{h}^1-f_\text{h}^2)^{2}}{(f_\text{h}^1+f_\text{h}^2)^{2}} \; ,
\label{Eq_Ah}
\end{equation}
where $f_\text{h}^1$ and $f_\text{h}^2$ are the hadronic energy fractions of two subjets respectively, as defined in Eq.~\eqref{hcaleq}.

\item  We denote the charge radius of the jet $J$ by $r_{C}$, and define it by 

\begin{equation}
r_{C} \ \equiv \ \frac{1}{d_0}
\sum_{k \in \text{tracks}} q^{(k)} \:  p_{T}^{(k)} \  \Delta R_{k J}    \; , \quad \text{where} \quad
d_0 = \sum_{k} p_{T}^{(k)}	 \; ,
\end{equation}
$q^{(k)}$ and $p_{T}^{(k)}$ are the charge and the transverse momentum of the $k$-th track inside the jet $J$
respectively,  $\Delta R_{k J}$ is the angular distance of the track from the jet axis.

\item  We estimate the neutral fraction of nonhadronic energy of the jet $J$, denoted by $f^N_{1-h}$ and defined as

\begin{equation}
f^\text{N}_\text{1-h} \ \equiv \ \frac{1}{E_J \times f_{1-h} } \ \sum_{k \in \text{ECal}} \delta_{q^{(k)},0} \ E^{(k)}_J  \; ,
\label{Eq_fnG}
\end{equation}

where $\delta_{q^{(k)},0}$ ensures that only the constituents with zero charge contributes.

\item  We also use $N$-subjettiness variables~\cite{Thaler:2010trnsubjettiness},  $\tau_{N}$, where $N=1,2,3, \dots $,  defined as 
\begin{equation}
\tau_{N} \ \equiv \ \frac{1}{R \: d_0 } \ 
\sum_{k} p_{T}^{(k)} \ \text{min}(\Delta R_{1,k},\Delta R_{2,k},...,\Delta R_{N,k})  \; ,
\end{equation}
 where $\Delta R_{i,k}$ is the distance of $k$-th constituent from the $i$-th axis and $R$ is the jet radius; we use axes of the exclusive $k_T$ subjets as the seed axes, and only do one pass at minimization.
\end{itemize}

\subsection{Identify four-vectors for the $e$-candidate and the $b$-candidate}
\label{sec_Ve}

The purpose of this subsection is to identify the four-vector corresponding to the electron from the top quark decay. The first step towards this aim is to identify the subjet containing an electron. We do this by keeping track of the distribution of energy in different parts of the calorimeter for the subjets, obtained after SD grooming, at various levels of declustering.   
To be specific, given any subjet represented  by $j$, we use $f_\text{h}^j$, defined in Eq.~\eqref{hcaleq}.

The algorithm we adopt in order to find the electron is as follows:

\begin{itemize}
	\item The subjet with lower $f_\text{h}^j$,  is identified to be the one most likely to contain an electron\footnote{This assignment has been checked, using generator-level information, to be correct for more than $85 \%$ of times.}. We refer the corresponding subjet by $\tilde{j}$.

	\item The hardest track in $\tilde{j}$ is denoted by  $\mathcal{T}$. In particular, the pseudorapidity and the azimuthal angle of $\mathcal{T}$ are referred to as $\eta_e$ and $\phi_e$.

	\item Constituents of $\tilde{j}$ are clustered further to find two exclusive $k_T$ subjets. Two exclusive $k_T$ subjets have been used as the electron is likely to participate only in the last stage of clustering in jet formation with the other particles, coming mostly from underlying event or neighboring $b$ quark.  Among these, the energy of the subjet containing $\mathcal{T}$ is recorded as $E_e$. 

	\item  The four-vector of  $e$ candidate, denoted by $p_e$,  is defined as
	\begin{equation}
 	p_e \  \equiv  \ \left\{  E_e,  E_{e} \sin \eta_{e} \cos \phi_{e} , E_{e} \sin \eta_{e} \sin \phi_{e}, 	E_{e} \cos \eta_{e} 
\right\}   \; ,
	\label{Eq_pe}
	\end{equation}

	\item The four-vector of the $b$ candidate (denoted by $p_b$) is defined using the four-vector of the full jet ($p_J$) and $p_e$ as follows : 
	\begin{equation}
	p_b \ \equiv \ p_J - p_e \; .
	\label{Eq_pb}
	\end{equation}
\end{itemize}

\subsection{Reconstruct approximate Neutrino momentum}
\label{V_nu}

In the limit the $W$ boson from the top quark decay is boosted, all $W$ boson decay products including the neutrinos are expected to be highly collimated. Starting with a high $p_T$ jet, where a candidate for an electron is identified, one, therefore, should be able to reconstruct the neutrino energy approximately, with the approximation getting better and better in the limit of high boost. We begin this subsection with three following central assumptions: 
\begin{itemize}
	\item The four-momentum of the electron candidate inside the jet  has already been identified, using the method described in the previous subsection.  We neglect electron mass throughout this work. Designating the direction of the electron three-vector  by the unit vector $\hat{e}$, we therefore have  
	\begin{equation}
	 p_e  \ \equiv  \ E_e \left( 1, \hat{e} \right) \; , \qquad \text{ with }     \hat{e}^2 = 1  \;.
	\end{equation}
	\item The four-momentum of the $b$ candidate identified in the previous subsection represents the momentum of the $b$ quark from the top quark decay.  Consequently, we will have non-negligible mass for the $b$ candidate. In this subsection  and later in this work we refer to the mass of the $b$ candidate by $m_b$.  
	\item Without loss of generalization, we decompose neutrino three-momentum vector, namely $\vec{p}_{\nu} $, into  $p^{\parallel}$, the component collinear to the direction of the $e$-candidate, and $ p^{\perp}$,  the component in the plane transverse to the direction of the $e$-candidate. More specifically,  
	\begin{equation}
	p_\nu  \ \equiv  \ E_\nu \left( 1, \frac{\vec{p}_{\nu}} {E_\nu}\right) \; , \quad \text{ with } 
	\vec{p}_{\nu} \ \equiv  \  p_{\parallel} \: \hat{e}  + \vec{p}_{\perp}
	\quad   \text{where}  \quad  
	\hat{e} \cdot \vec{p}_{\perp} = 0 \; . 
	\end{equation}
	The key assumption is that the neutrino is mostly collimated with the $e$ candidate. More specifically, 
	\begin{equation}
	r  \  \equiv \  \frac { \left| \vec{p}_{\perp} \right| } {p_{\parallel}}  
		\ = \ \frac { p_{\perp}  } {p_{\parallel}}  \ \ll \ 1  \; .
	\end{equation}
	This assumption allows us to simplify and expand neutrino energy in a power series in $r$, which would be crucial later. 
	\begin{equation}
	E_{\nu} \ = \ \left( \sqrt{ 1 + r^2 } \right)  p^{\parallel} \ \simeq  \  \left( 1 + \frac{1}{2} r^2 \right) p^{\parallel} + \mathcal{O}\left( r^4\right)
	\end{equation}
\end{itemize}

We show in discussion below, the set of  assumptions itemized above is sufficient in order to derive the neutrino energy in the high boost regime. The kinematics of $W$ boson decay forces the following relation: 
\begin{equation}\label{eq:kin-w}
m_{W}^2 \ = \  \left( p_e + p_{\nu}\right)^2 \ = \  
2  E_e E_\nu \left( 1 -   \frac{p_{\parallel}}{E_\nu} \right)  \ = \ 
2  E_e E_\nu \left( 1 - \frac{1}{\sqrt{ 1 + r^2 }} \right)  \ \simeq \
r^2  E_e E_\nu
\end{equation}
Similarly, top quark decay kinematics renders an additional constraint.  
\begin{equation}\label{eq:kin-top}
\begin{split}
m_t^2 \ = \ \left( p_b + p_e + p_{\nu}\right)^2 \ = \  
m_{W}^2 +  m_{b}^2 +  2 p_{b} \cdot p_{e} + 2 E_{b}E_{\nu} - 2\vec{p}_{b} \cdot \vec{p}_{\nu} \\
=  m_{W}^2 +  m_{b}^2 +  2 p_{b} \cdot p_{e}  + 
2 E_{\nu} \left\{ E_b - \frac{1}{\sqrt{1+r^2}} \left(  \vec{p}_{b} \cdot  \hat{e} + r \: \vec{p}_{b}  \cdot \frac{\vec{p}_{\perp}} {p_\perp} \right)  \right\} \\ 
\simeq  \Delta^2 + 2 E_\nu \left( E_b -  \vec{p}_b \cdot \hat{e} 
- r  \: \vec{p}_b \cdot \frac{\vec{p}_{\perp}}{p_{\perp}} + \mathcal{O}\left( r^2\right) \right)     \; , 
\end{split}
\end{equation}
where $\Delta^2 \equiv \left(m_{W}^2 +  m_{b}^2 +  2 p_{b} \cdot p_{e} \right) $ is a measured quantity. Eq.~\eqref{eq:kin-w} and Eq.~\eqref{eq:kin-top} can be solved simultaneously to yield expression for $r$, and therefore $E_\nu$.
\begin{equation}\label{eq:nu_soln}
\begin{split}
r^2 \ & \simeq \   \frac{2 m_{W}^2}{m_t^2 - \Delta^2}  \frac{ \left( E_b - p_{b} \cdot \hat{e} \right)}{E_e}  \\
E_\nu \ & \simeq \ \frac{1}{2} \frac{m_t^2 - \Delta^2}{\left( E_b - p_{b} \cdot \hat{e} \right)} 
\end{split}
\end{equation} 
Instead of directly using  $E_\nu$ as determined from Eq.~\eqref{eq:nu_soln}, we rather employ dimensionless quantities defined as follows:  
\begin{align}
Z_{b} \ & \equiv \  \frac{E_b}{ E_e + E_\nu} 
\label{eq:b_split}  \\
\Theta_{b/e}  \ & \equiv \  \frac{E_e \left( m_t^2 - \Delta^2 \right) }{E_b m_W^2} 
 \label{eq:rel_angle} 
\end{align}
Within the ansatz that the observed jet arises from a top quark decay and contains a massless neutrino collimated with the electron, both these variables have simpler interpretation. $Z_b$ represents the ratio of the relative fraction of top energy carried by the $b$ candidate with respect to that carried by the $W$ boson candidate, and $\Theta_{b/e} $  reduces to the ratio of the size of the opening angle between the reconstructed neutrino and the $b$ candidate with respect to the angle between  the neutrino and the $e$ candidate.
\begin{equation}
Z_b \   \rightarrow \ \frac{E_b}{E_W}   \qquad \text{and} \qquad  
\Theta_{b/e}   \  \rightarrow \ 
\frac{1 - \left( \vec{p}_\nu \cdot  \vec{p}_b \right)/ E_\nu E_b  }{1 - \left( \vec{p}_\nu \cdot  \vec{p}_e \right)/ E_\nu E_e }
\ \simeq \ 
\frac{1 - \cos \theta_{\nu b} }  {1 - \cos \theta_{\nu e} }   \; , 
\end{equation}
where $\theta_{\nu b}$ and $\theta_{\nu e}$ are the opening angles of the reconstructed neutrino from the direction of the $b$ candidate and the $e$ candidate respectively. Note that the approximate sign in the Eq.\eqref{eq:rel_angle} arises due to the approximation $\left| \vec{p}_b \right| \simeq E_b$. 
The set of these two variables, $Z_b$ and $\Theta_{b/e}$, is referred as $\mathcal{V}_\nu$.

\subsection{Multivariate analysis and vetoing the background}
\label{Sec_BkgVeto}

The purpose of all the previous subsections in this methodology section is to calculate a bunch of variables given an input jet. As mentioned previously, we divide these variables in two sets. The variables in the set  $\mathcal{V}_e$ allows us to check whether the given jet may contain an energetic electron.  On the other hand, variables in $\mathcal{V}_\nu$ are only interpreted correctly if and only if the four-momenta corresponding to the electron candidate  and the $b$ candidate reconstruct $W$ and top correctly when an \emph{invisible} four-momentum corresponding to the neutrino and roughly collimated to the electron is added. 

Instead of treating all the variables on equal footing, we construct two BDT based multivariate discriminators, which separate electronic top jets (to be treated as signal jets) from QCD $b$ jets (to be treated as background jets). To be specific, let us define  
\begin{equation}
\begin{split}
\mathcal{B}_{e}^{t/b} \ \equiv \  \text{A BDT to discriminate } t \text{ from } b \text{ using variables in } \mathcal{V}_e  \; ,  \\ 
\mathcal{B}_{\nu}^{t/b} \ \equiv \  \text{A BDT to discriminate } t \text{ from } b \text{ using variables in } \mathcal{V}_\nu \; .
\end{split}
\label{Eq_BDT}
\end{equation}
Upon optimizing on the samples of electronic top jets and QCD $b$ jets, these BDTs learn to give different responses for top jets than to  QCD $b$ jets. We rescale the BDT responses such that each of these now range in $\{ -1, +1\}$; as a result, QCD $b$ jets mostly get characterized by values close to $-1$, whereas electronic top jets lie close to $+1$.  Denoting the responses by $r_e$ and $r_\nu$, and defined by 
\begin{equation}
\begin{split}
r_e \  \equiv \  \text{response of } \mathcal{B}_{e}^{t/b} \text{ in the range }   \{ -1, +1\}  \; ,   \\
r_\nu \  \equiv \  \text{response of } \mathcal{B}_{\nu}^{t/b} \text{ in the range }   \{ -1, +1\} \; , 
\end{split}
\label{Eq_R}
\end{equation}
we construct a plane of responses, where any jet is represented by a point.  We show the distributions of BDT responses for electronic top jets and QCD $b$ jets later in the Sec.~\ref{subsec:results_mva}. However, it is easy to visualize that, by construction, all  QCD $b$ jets dominantly occupy locations near $ (-1, -1)$ corner of the plane, whereas electronic top jets populate the region around the  corner corresponding to coordinates $(+ 1, +1)$. 

It is therefore a straightforward exercise to construct a tagger for electronic top jets, which at the same time can find anomalies that can be considered as outliers as far as QCD $b$ jets as well as  electronic top jets are concerned. 
\begin{itemize}
	\item Find a zone boundary around the corner  $ (-1, -1)$ of the response plane that demarcates a zone containing within itself a \emph{pre-assigned} fraction of QCD $b$ jets.  We block this zone so that any jet characterized by responses falling within this zone are ``vetoed". 
	\item In principle, we can tag any jet that fails the veto criteria to be an electronic top jet, which yields a large tagging efficiency. However, we rather find another zone around the corner  $ (+1, +1)$ of the response plane which contains within itself again a  \emph{pre-assigned} fraction of  electronic top jets. We designate this zone to be the signal zone and any jet characterized by coordinates in the signal zones would be tagged as an electronic top jet. 
	\item Jets not belonging to either the veto zone or the signal zone are clearly outliers. We term these as anomalous jets.    
\end{itemize}

\section{Simulations Details}
\label{sec:sample}

We demonstrate the potency of the methodology proposed in the last section using simulated samples. In this section, we lay out clearly the simulation details and the details of generated samples we use for benchmarks.
\begin{itemize}
	\item We generate all the Monte Carlo (MC) samples at partonic level using  {\textsc{MadGraph}\xspace}5 (MadGraph5\textunderscore aMC$@$NLO~V5~2.6.2)~\cite{Alwall:2014hca} at leading order in perturbative QCD with {\textsc{NNPDF2.3LO}\xspace}~\cite{Ball:2014uwa} parton distribution function at centre-of-mass energy of $13$~TeV. 
	The renormalization scale, at which the strong coupling constant ($\alpha_S$) is evaluated, is taken to be $Z$ boson mass; the same value is chosen for the factorization scale also.
	We employ {\textsc{pythia}\xspace}8.230~\cite{Sjostrand:2014zea}   for showering and hadronization.
	The renormalization scales for initial- and final-state radiation are taken to be the same as the renormalization scale at matrix element level, i.e. $Z$ boson mass.
	 Additionally, we use {\textsc{4C}\xspace} tune \cite{Sjostrand:2006za} to simulate the relevant busy hadronic collider environment. 
	
	\item  In order to provide a semi-realistic environment for high energy collisions, we use {\textsc{Delphes}\xspace} 3.4.1 software package~\cite{deFavereau:2013fsa} with CMS geometry~\cite{CMS_ex}, where  stable particles from {\textsc{pythia}\xspace} are converted into detector objects such as energy deposits in calorimeter cells and tracks. The particle-flow (PF) algorithm~\cite{Sirunyan:2017ulk}, as implemented in {\textsc{Delphes}\xspace} uses these detector elements to construct particle-flow candidates, namely muons, electrons, photons, charged and neutral hadrons. 
	
	\item We use anti-$k_T$ jet clustering algorithm in   {\textsc{Fastjet}\xspace}~\cite{Cacciari:2011ma} to cluster the particle-flow candidates into jets. In particular, we use $R = 0.8$ with $p_{T_{\text{min}}} = 500~\text{GeV}$.  The hardest  jet (in $p_T$)  in the event is used in our analysis.  

\end{itemize}
We do not simulate effects from  pileup. Since  we use a large threshold for the transverse momentum of  jets to be considered, and also perform soft drop grooming, we expect the effect of pileup  to be minimal.

We benchmark the performance of the proposed tagger by considering a number of labeled samples of jets, which are all generated via the procedures described above. 
\begin{itemize}
	\item electronic top jets: we generate events $pp \rightarrow t \bar{t}$, where each top quark decays via $t \rightarrow b e \nu_e$. During production we impose a $p_T$-cut of $500$~GeV on the $b$ quarks coming from top quark decay, and a $p_T$-cut of 200 GeV on other quarks and gluons produced at the matrix element level. Because of the phase space cuts and specification of jet construction (listed above) we expect the leading jet to contain all the decay products of top quark. We use the same phase space cuts for other samples as well.  The leading jets in these events  are kept for further analyses. From now on we refer to these jets as $t(e)$ jets. 
	
	\item QCD $b$ jets: we generate events $pp \rightarrow b \bar{b}$ with the same phase space cuts as in top jets. Only the leading jet from each event is analyzed and we  refer to these as $b$ jets.  
	\item light flavor jets: we generate events $pp \rightarrow j j'$, where $j$ ($j'$) represents either a quark ($u$/$d$/$s$/$c$) or a gluon. From now on, we refer to the the leading jets in this sample of  events  simply as  $j$ jets.
	\item hadronic top jets: we generate events $pp \rightarrow t \bar{t}$, where each top quark decays via  $t \rightarrow b j j'$.  For this sample, we refer the leading jets as $t(h)$ jets.
	\item anomalous jets: we choose the minimal supersymmetric standard model (MSSM) as a benchmark to produce anomalous jets. Since we want a jet closely mimicking an electronic top jet, we use stop, nearly degenerate in mass with top.  To be specific, we generate events $pp \rightarrow \tilde{t} \bar{\tilde{t}}$ where the stop quark, $\tilde{t}$,  decays into $b e \nu_e \chi_0$, where the neutralino, $\chi_0$, represents the lightest supersymmetric particle. We choose the mass of $\tilde{t}$ to be $200$ GeV and that of $\chi_0$ to be $100$ GeV. We refer the leading jets for this event sample as $\tilde{t} (e)$ jets.
\end{itemize}
Before proceeding, let us emphasize once again that we do not intend to benchmark the proposed tagger for all possible anomalous jets. We rather show an example of anomalous jets which can be identified with reasonable acceptance as a proof of principle.

\section{Results}
\label{sec:results}
\begin{figure}[hbtp]
\begin{center}
    \includegraphics[width=0.45\textwidth]{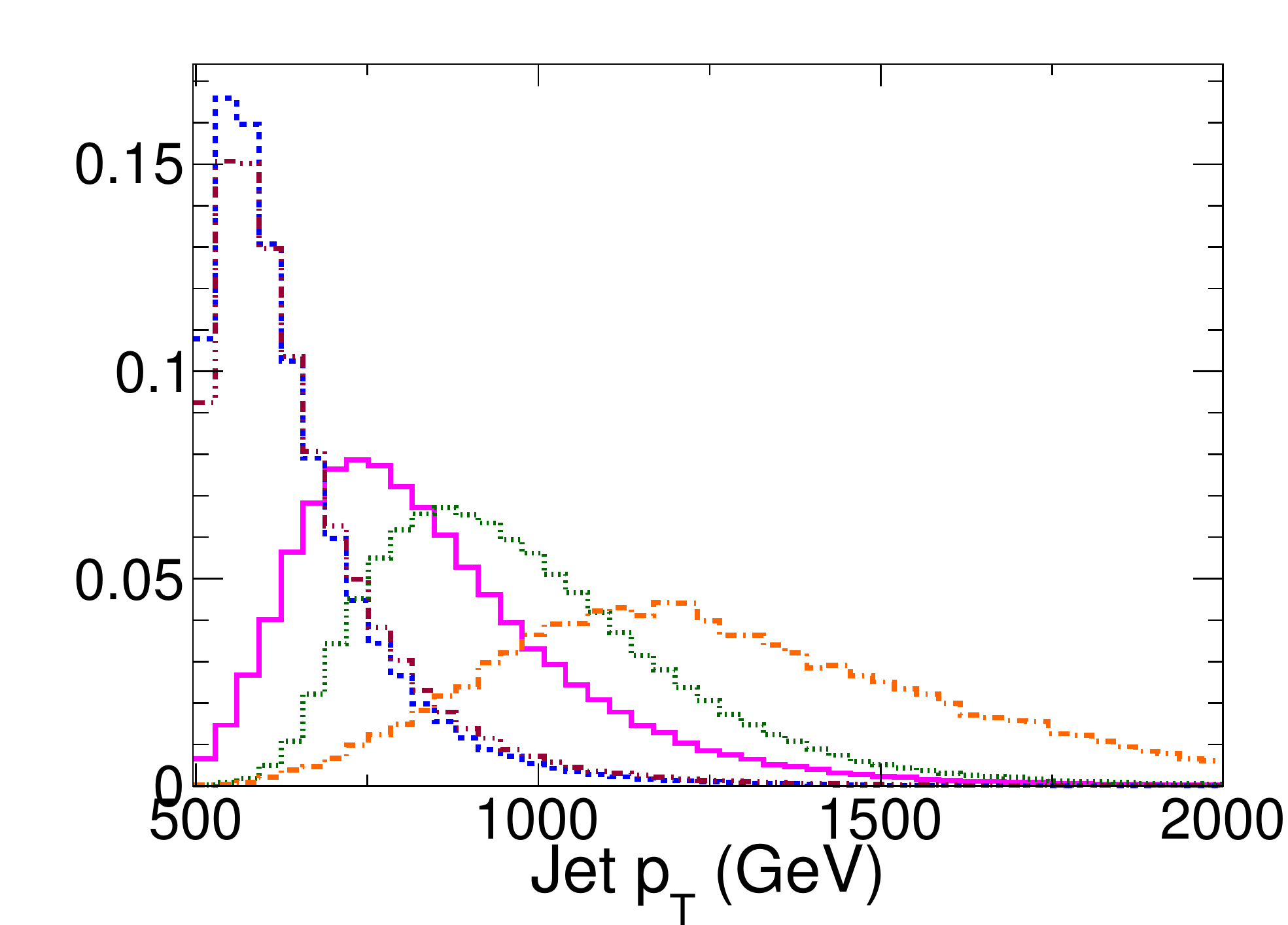}
    \includegraphics[width=0.45\textwidth]{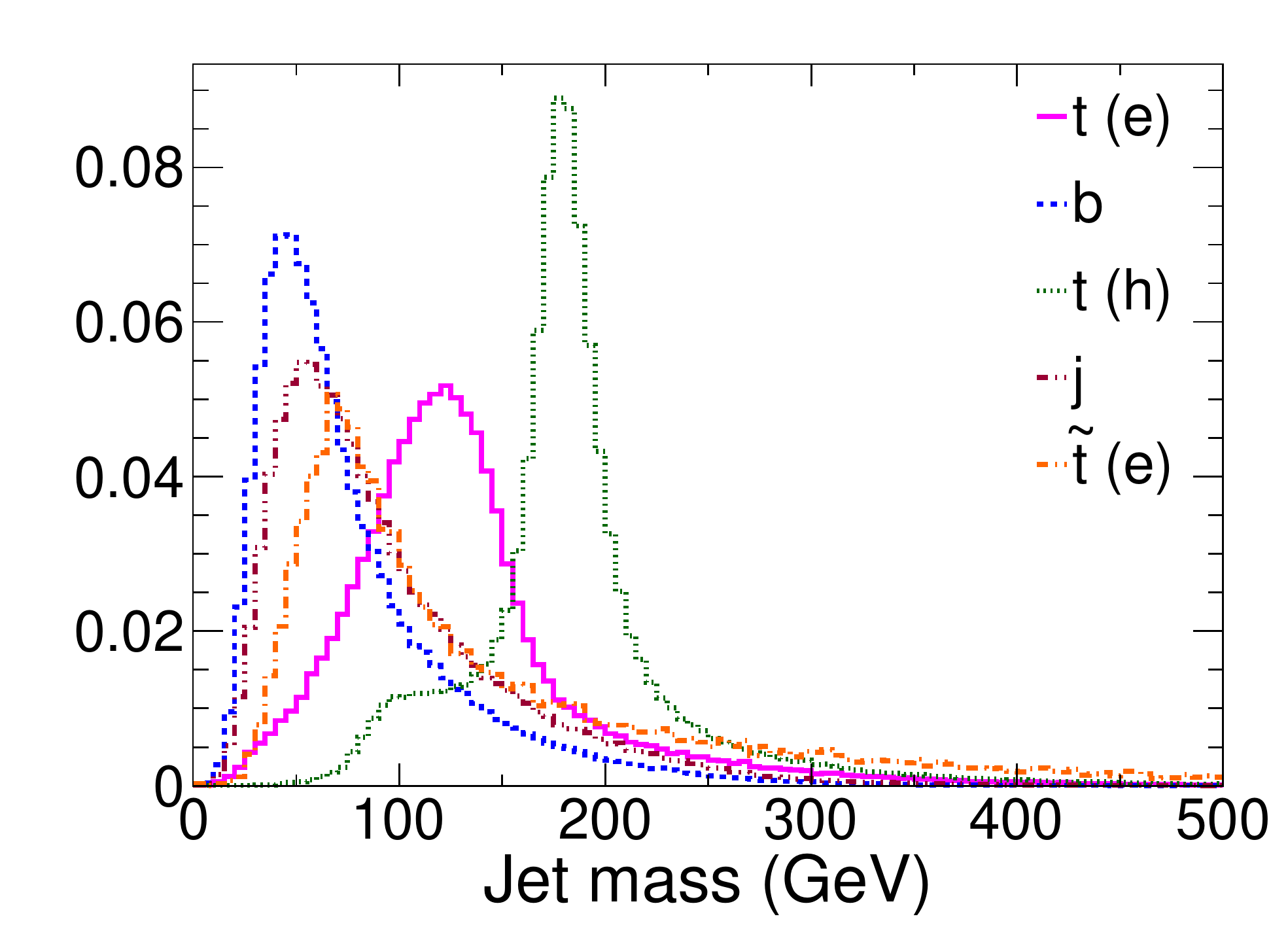}
	\caption{Distribution of $p_{T}$ (left), and mass (right) of the ungroomed jet for all the event samples.} 
    \label{fig:variables_ptm}
\end{center}
\end{figure}

In this section we summarize the results of our studies using the simulated samples described in the previous sections. Note that as mentioned before all jets we consider here are constructed using  anti-$k_T$ jet clustering algorithm with $p_T > 500~\text{GeV}$.  Before proceeding we show the phase space distributions of the ungroomed jet for each sample in Fig.~\ref{fig:variables_ptm}.  

For $b$ jet and light jet QCD samples, $p_T$ spectrum falls rapidly as expected; for top quark samples, as the  $p_T$ cut is put on the partons (b/c/s/u/d/g) during event generation, the  $p_T$ distribution of the jet peaks at a higher value.
As expected, for hadronically decaying top quark, the mass distribution peaks around the top quark mass; for semileptonically decaying top quark, as the neutrino escapes the detector with a significant fraction of top quark energy, mass distribution has larger population between $W$ boson mass and top quark mass; for other samples, as expected, it's a falling distribution, other than a threshold because of the $p_T$ cut on the jet.

\subsection{Variables in $\mathcal{V}_e$} 
\label{Sec_Results_Ve}

Out of all the variables defined in Sec.~\ref{Sec_Ve}, we use the following $6$ variables: 
\begin{equation}
\mathcal{V}_e \ \equiv \ \left\{   f_\text{1-h}, {A}_h,  f^\text{N}_\text{1-h} , \tau_{21} \equiv \frac{\tau_2}{\tau_1}, r_C, m_{\text{SD}} \right\} \; .
\end{equation}

\begin{figure}[hbtp]
	\centering
	\includegraphics[width=0.45\textwidth]{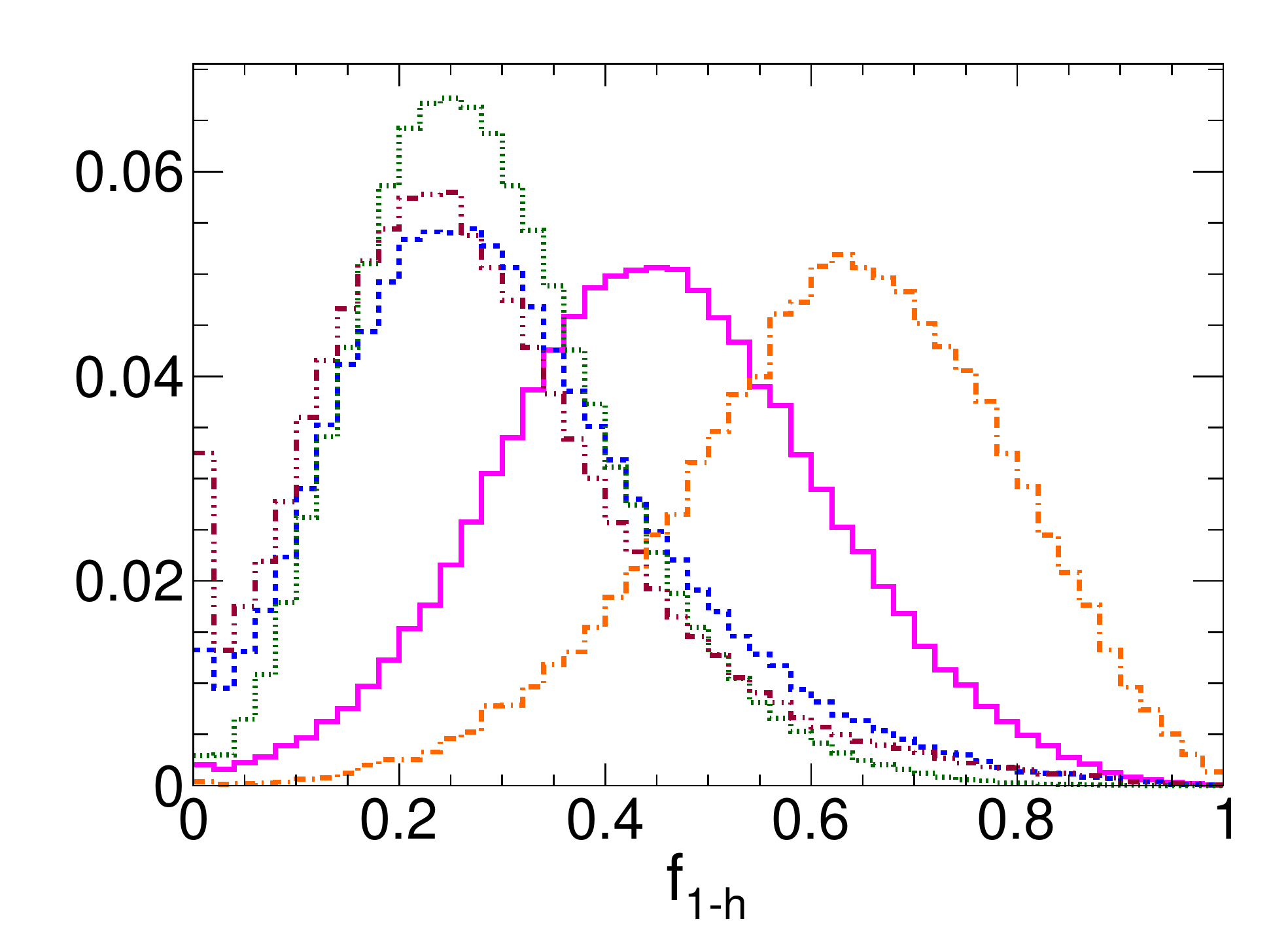}
	\includegraphics[width=0.45\textwidth]{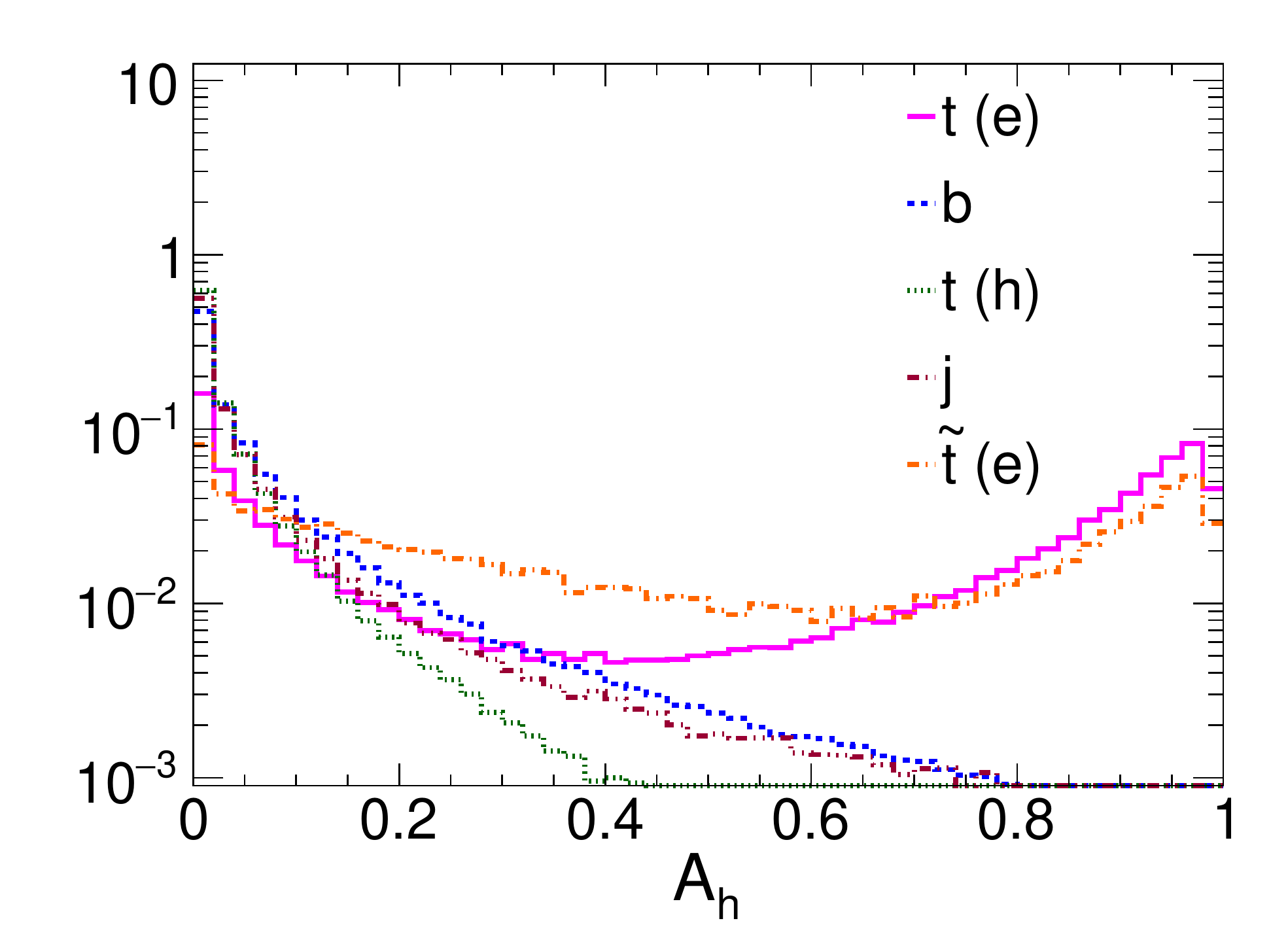}
	\includegraphics[width=0.45\textwidth]{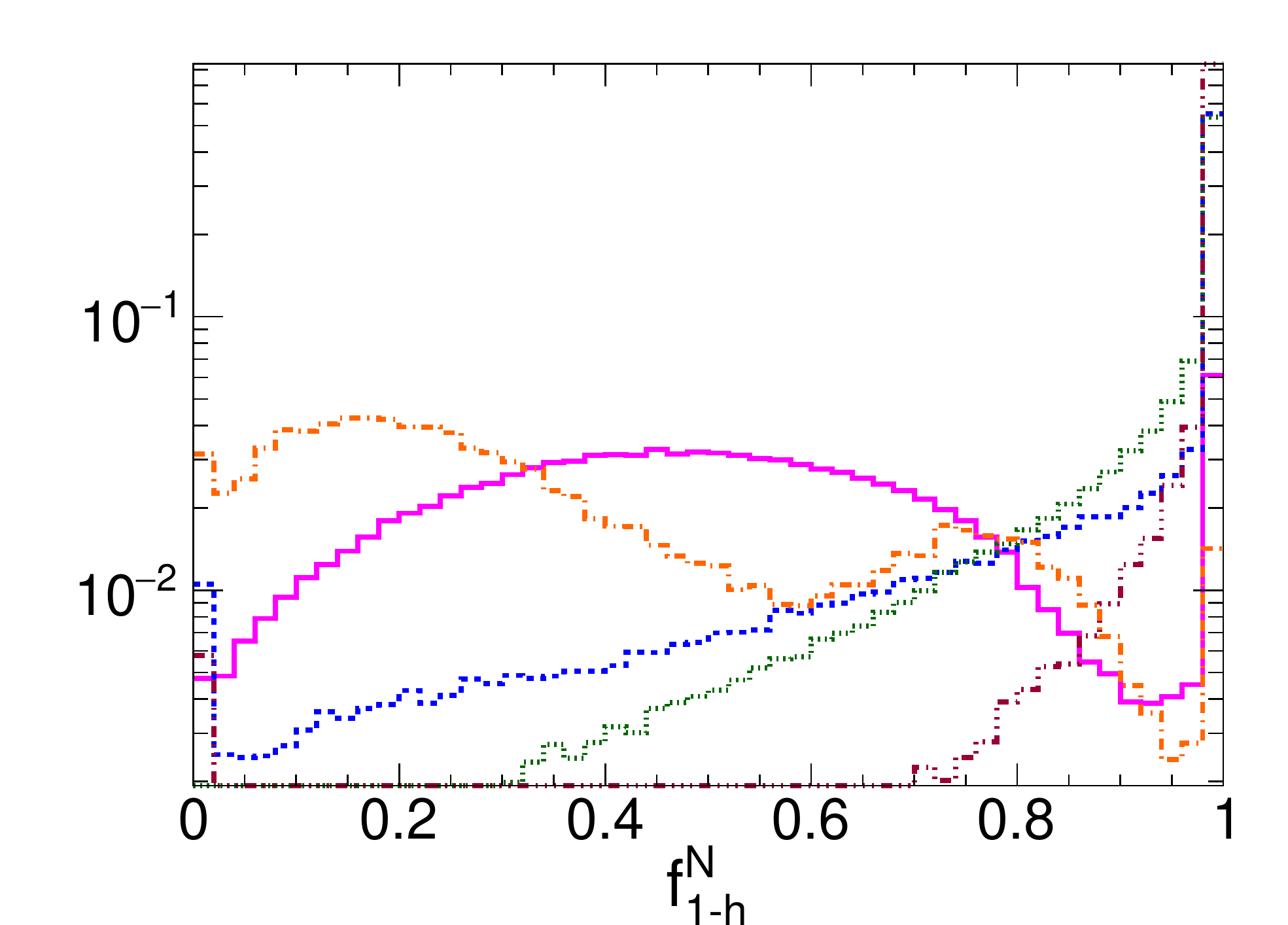}
	\includegraphics[width=0.45\textwidth]{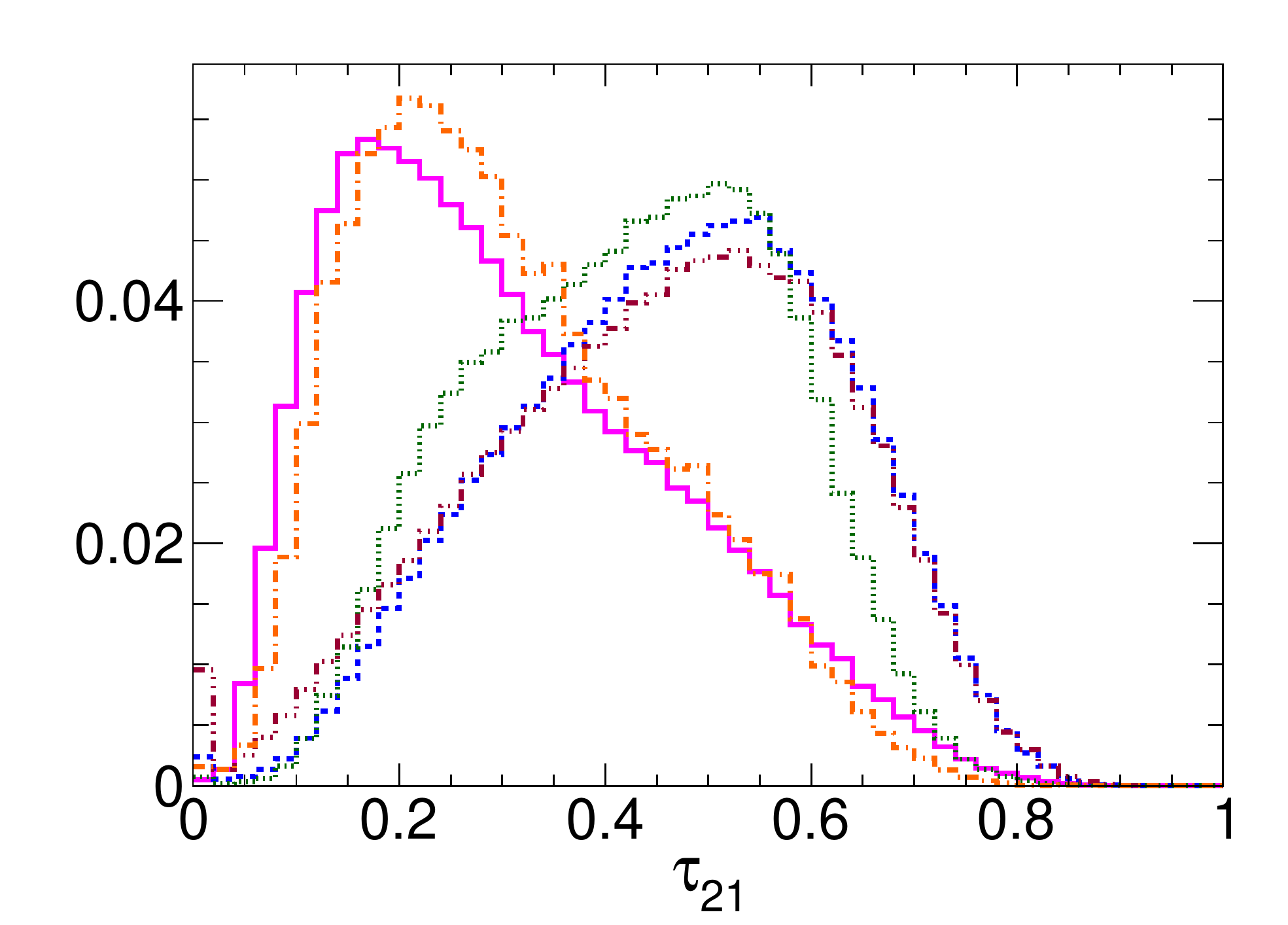}
	\includegraphics[width=0.45\textwidth]{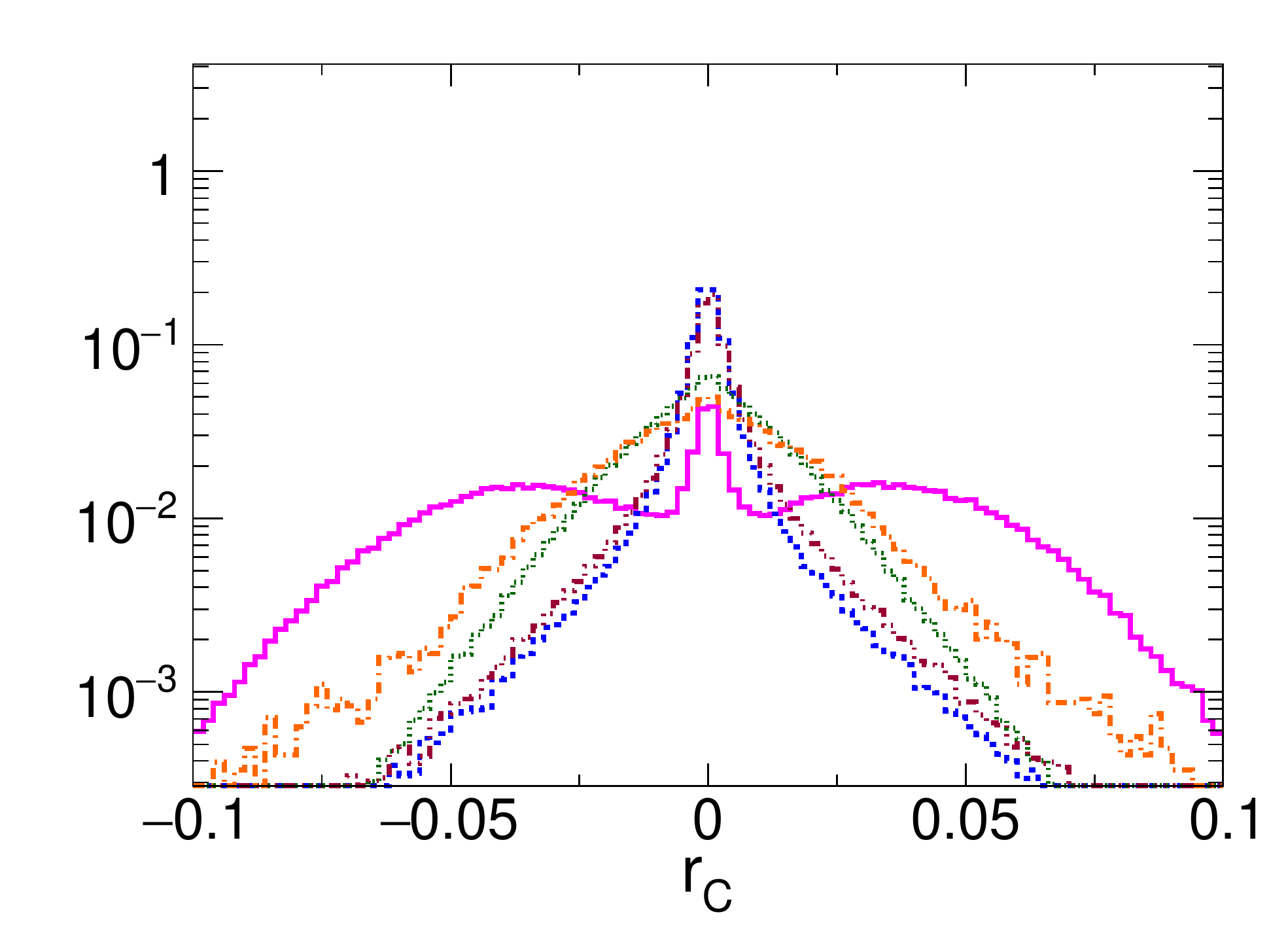}
	\includegraphics[width=0.45\textwidth]{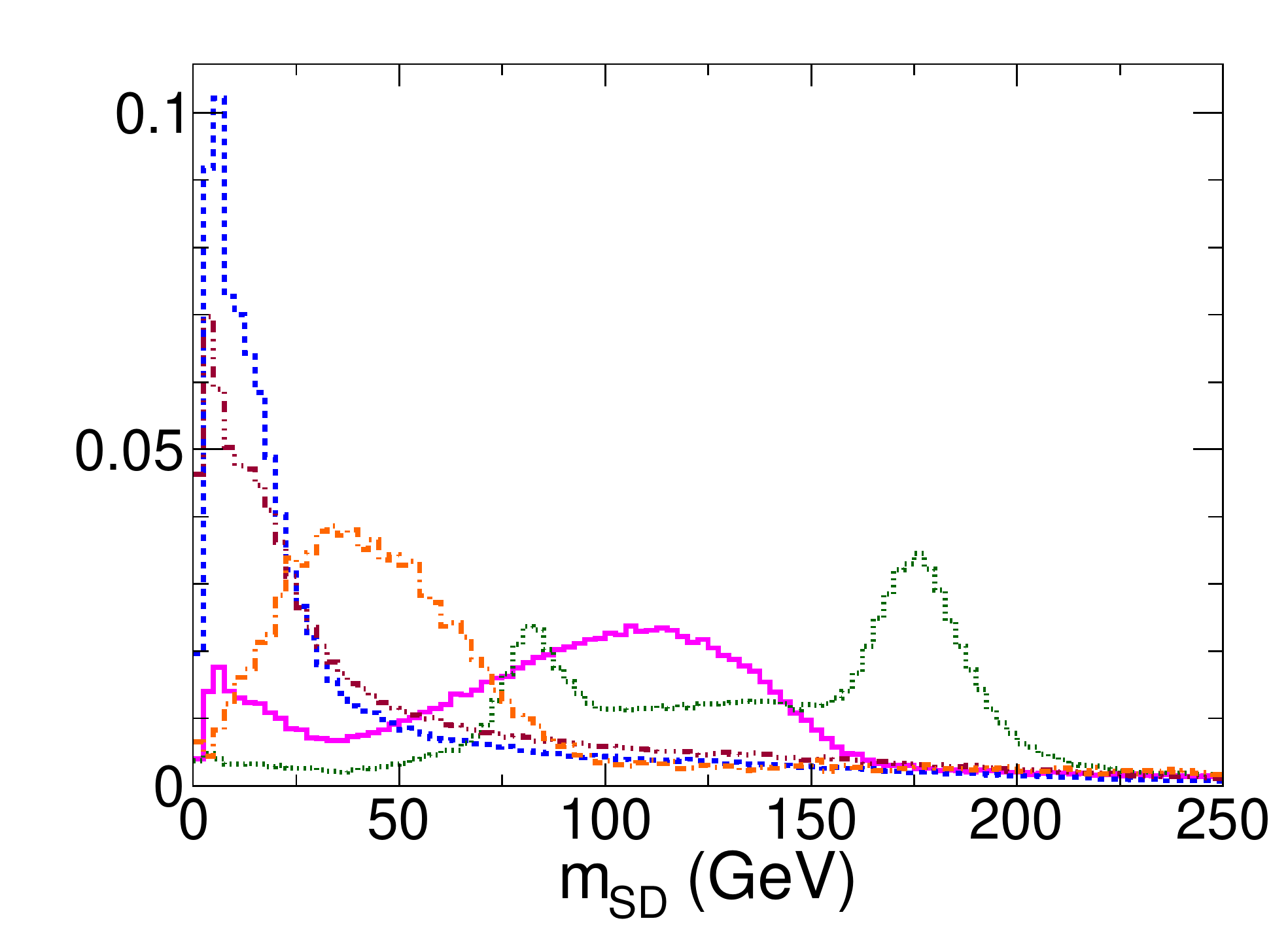}
	\caption{Distribution of $f_\text{1-h}$ (top left), $A_h$ (top right),  $f_\text{1-h}^N$ (middle left),  $\tau_{21}$ (middle right), $r_C$ (bottom left), and $m_{SD}$ (bottom right) of the jet for different event samples considered.}
	\label{fig:variables_1}
\end{figure}

We show the distributions of these variables for different jet samples in Fig.~\ref{fig:variables_1}. Out of the six variables, the first two are rather straightforward to estimate.  In order to calculate the nonhadronic energy fraction, one simply needs the energy deposited in HCal cells, and subtract it from the total energy of the jet.  Note that since we use particle-flow candidates to construct jets, jet constituents are classified as charged and neutral hadrons, photons, electrons, and muons. In our case, we simply add energies of all charged and neutral hadrons within the jet to find total energy deposited in the HCal. Similarly, we add energies of all electron and photon particle-flow candidates to estimate the ECal energy.  
In the top left panel we show the distributions of the nonhadronic energy fraction or $f_\text{1-h}$. The $t(e)$ jets rich in energetic electrons from top quark decays show up with significant larger  $f_\text{1-h}$ as expected.  In case of other jets, mostly consisting of  hadrons,  one typically expects $\sim 30\%$ energy in the original jet carried by photons from $\pi^{0}$ production in hadronization. There could also be a small fraction of energy in the ECal because of semileptonic decays of heavy flavor quarks or  even  some  energy deposit from nuclear interaction of the hadrons in the ECal. Still, the fraction of energy deposit in the ECal is smaller as compared to that of electronic top jets.  
As shown in  the top right panel of  Fig.~\ref{fig:variables_1},  the hadronic asymmetry among subjets  or ${A}_h$  is clearly one of the most powerful variables that can tell apart a $t(e)$ jet from the background jets. The asymmetry is maximized in case of $t(e)$ jets, which are largely characterized by one subjet initiated  by a $b$ quark and the other by an electron. In case of light  flavor jets or even for $b$ jets or  $t(h)$ jets, we expect both the subjets to be initiated by quarks or gluons, and, therefore, rather symmetric~\footnote{We have also studied an alternative definition of $A_h$, where we use the $b$ candidate and the $e$ candidate, as described in Eqs.~\eqref{Eq_pe} and~\eqref{Eq_pb} respectively, in place of two subjets $j_1$ and $j_2$ in Eq.~\ref{Eq_Ah}. However, since we do not find any improvement in the performance, we employ subjets $j_1$ and $j_2$ in order to define $A_h$ for simplicity.}.
Estimating the neutral fraction of nonhadronic energy or $f_\text{1-h}^{N}$ is slightly nontrivial. One way to estimate it would involve finding energy deposits in ECal cells that don't correspond to any track matched to it. Since we already use particle-flow elements, we simply need to find the total energy carried by the photon components of the particle-flow constituents.  In the left figure of the middle panel we show the distribution of  $f_\text{1-h}^{N}$  for different jet types under consideration.  As argued before, almost all of electromagnetic energy of light flavor jets are because of photons from $\pi^0$ decays, which gets demonstrated in the plot neatly where one finds a sharp peak for $f_\text{1-h}^{N}$ near $1$. Note that $b$ jets or $t(h)$ jets are also characterized in a similar manner, except of little more flattening for small   $f_\text{1-h}^{N}$ signifying semileptonic decays of heavy flavor quarks. Not surprisingly, $t(e)$ jets show rather moderate amount of energy in photons, which are largely due to bremsstrahlung radiations in case of energetic electrons from top quark decay,  while travelling through the tracker and the calorimeter.  

Jets from electronic decay of the top quarks result in two distinct hard subjets, one of which corresponds to a subjet initiated by the $b$ quark and the other one because of an energetic electron. The $N$-subjettiness ratio,  in particular, $\tau_{21}$ is especially suitable to find these objects.  Since $t(e)$ jets are generically characterized by $2$ subjets, one finds $\tau_2 \ll \tau_1$, or in other words,  $\tau_{21} \ll 1$ as can be seen in the middle right panel of figure~\ref{fig:variables_1}. For all other jets we tend to get comparatively  larger values of $\tau_{21} $. Note that some of the times $t(h)$ jets end up loosing parts of the decay products from top and end up having 2 hard subjets, and hence small $\tau_{21}$. The slight shift of  $t(h)$ to the left as compared to $b$ jets signify these cases.  Another manifestation of the similar physics can be seen in the bottom  left panel of figure~\ref{fig:variables_1}, where we plot the charge radius or $r_C$ of for the jets.  The winglike feature observed for $t(e)$ jets  signify existence of hard charged constituents of the jets at the periphery.  Finally we plot the soft drop mass ($m_\text{SD}$) for jets in the bottom right panel. Not surprisingly,  the distribution for $t(h)$ jets peaks at around the top quark mass when the top is fully captured inside the fat jet, and also around $W$ boson mass for partial reconstruction. The  $m_\text{SD}$ distribution for QCD jets are well studied and well understood~\cite{Marzani:2017mva,Bendavid:2018nar}, whereas those for $t(e)$ jets and $\tilde{t}(e)$ jets are characterized by missing masses because of missing energies due to neutrinos and neutralinos respectively. 

Not surprisingly,  the set of anomalous jets we consider, namely $\tilde{t}(e)$ jets, almost always lies in between electronic top jets and hadronic top jets, and  only $f_\text{1-h}$ makes these outliers. This is easy to understand. Since the neutralino carries away a large chunk of energy (and mass), it decreases total visible jet energy. Energy carried away by the electron from stop quark decay is also smaller than that of the electron from top quark decay, but that change is relatively small. This is part of the reason why $\tilde{t}(e)$ jets are characterized by  larger  $f_\text{1-h}$  than jets of any other type.  Similarly because of larger missing mass,  $\tilde{t} (e)$ jets contain smaller $m_\text{SD}$ than $t(e)$ jets. Since $\tilde{t}(e)$ jets contain mostly two hard subjets, one initiated by the $b$ quark and the other by the electron, the ${A}_h$ and $\tau_{21}$ plots are, in fact,  identical with $t(e)$ jets.

\subsection{Variables in $\mathcal{V}_{\nu}$} 
\label{Sec_Results_Vnu}
\begin{figure}[hbtp]
	\begin{center}
		\includegraphics[width=0.45\textwidth]{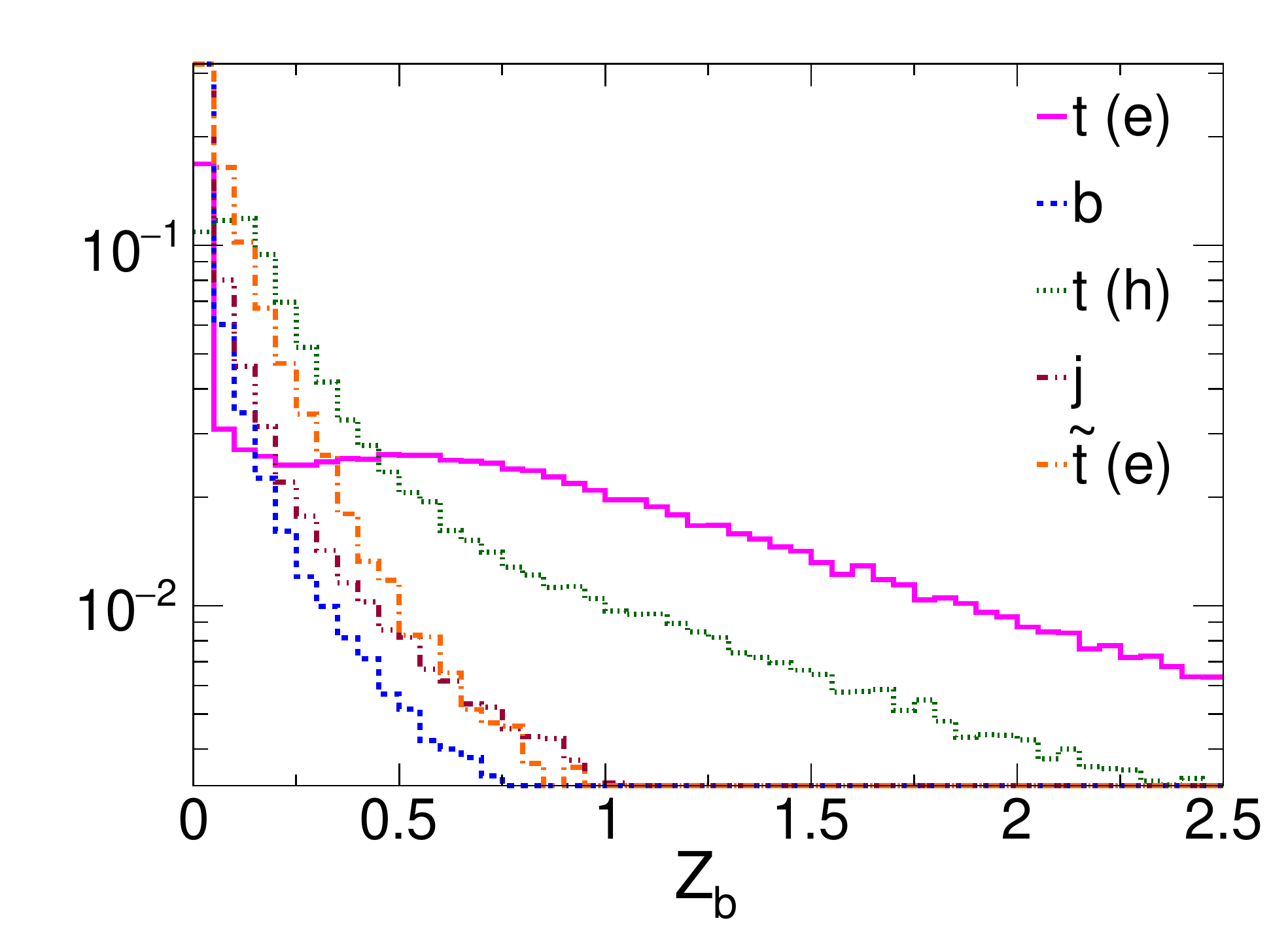}
		\includegraphics[width=0.45\textwidth]{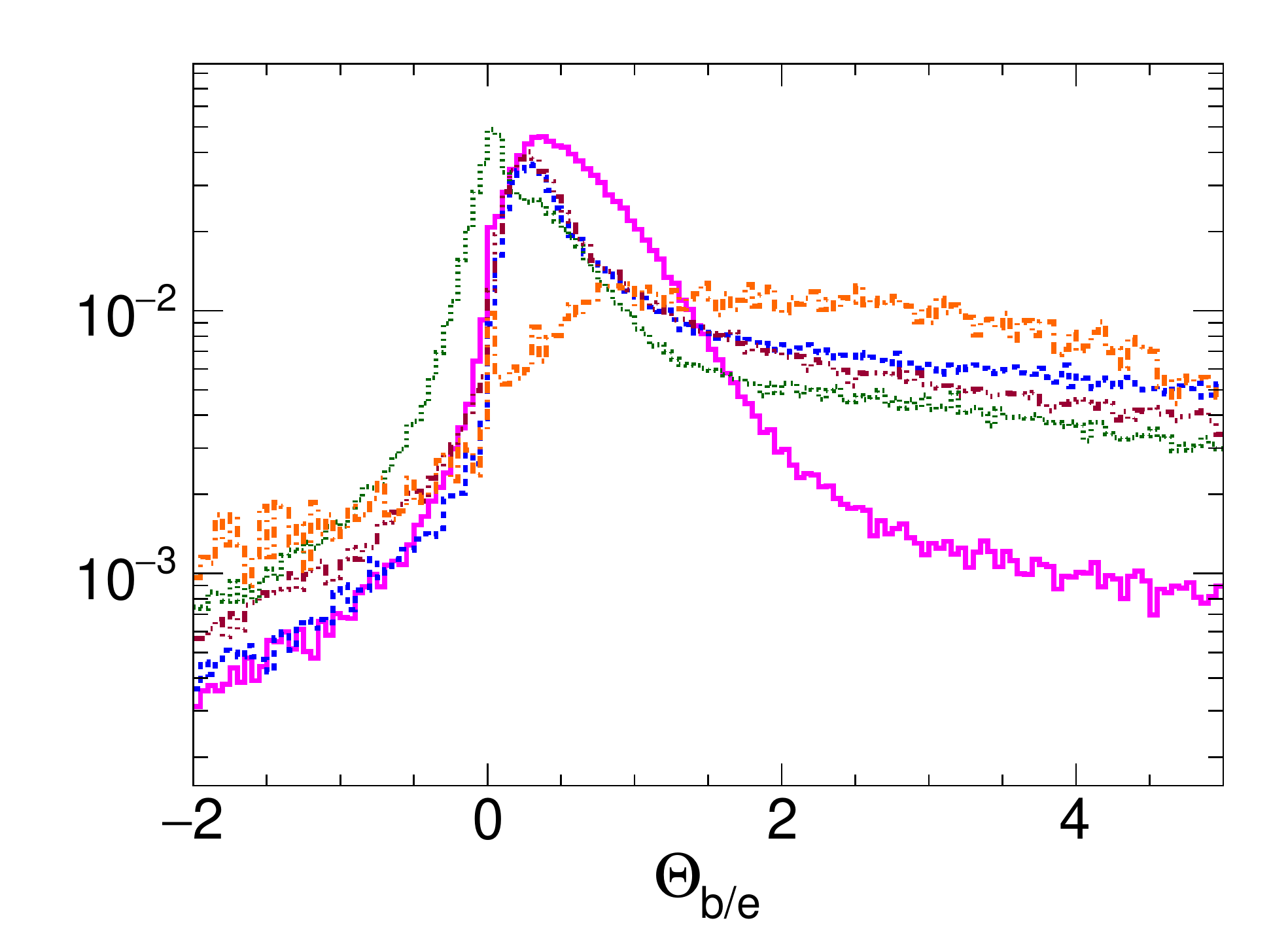}
		\caption{Distribution of $Z_{b}$ (left), and $\Theta_{b/e}$ (right) for different jet types.} 
		\label{fig:variables_3}
	\end{center}
\end{figure}

 As described in Sec.~\ref{V_nu},  the set of variables in $\mathcal{V}_{\nu}$ are given as: 
\begin{equation}
\mathcal{V}_\nu \ \equiv \ \left\{   Z_{b}, \Theta_{b/e} \right\} \; . 
\end{equation} 
We show the distributions of these variables in Fig.~\ref{fig:variables_3} for various types of jets under consideration. The physics discussion of these variables are already given  in Sec.~\ref{V_nu}. Note that these variables are constructed with the ansatz that if a collimated, massless missing four-vector is added to the $e$ candidate and the total jet, one reproduces the $W$ particle and the top particle respectively. Only $t(e)$ jets justify this ansatz and, therefore, show vastly different behavior than any other jets.   We expect $t(h)$ jets to satisfy this condition more often than $b$ jets and light flavor jets. Consequently, we find  curves representing $t(h)$ jets to lie in between $t(e)$ jets and $b$ jets. 

The stop jets considered here also fail the ansatz since the missing momentum is massive. The most important feature of these plots is that the distributions of $\tilde{t}$ jets are more closely described by $b$ jets than any of the top jets. This can be understood from the fact that both $b$ jets and $\tilde{t}(e)$ jets fail the ansatz (even though for vastly different reasons).   As we show later, this feature plays an important role in constructing the anomaly finder.

\subsection{Correlations of variables} 
\label{subsec:results_mva}
\begin{figure}[hbtp]
	\begin{center}
		\includegraphics[width=0.475\textwidth]{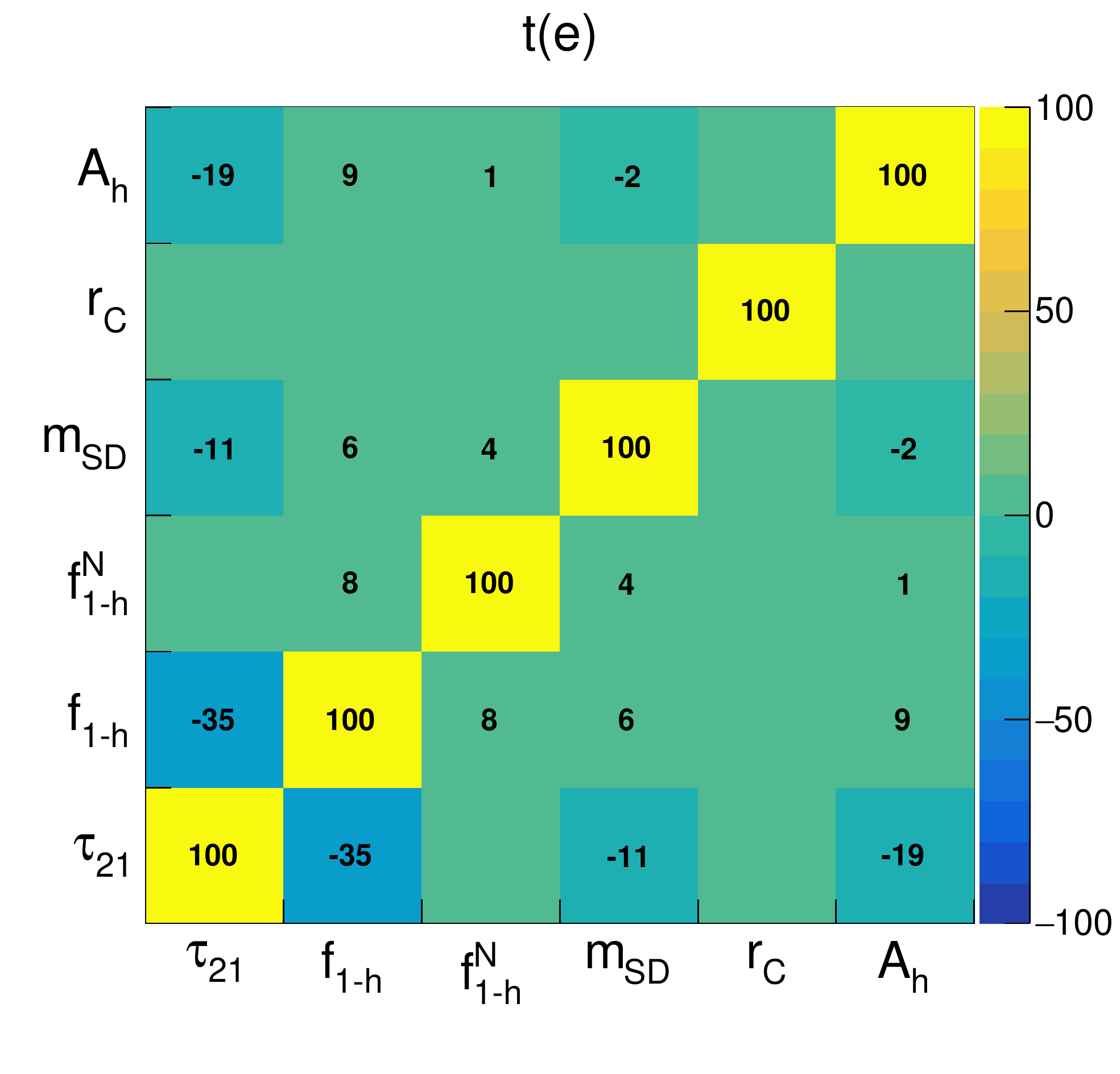}
		\includegraphics[width=0.475\textwidth]{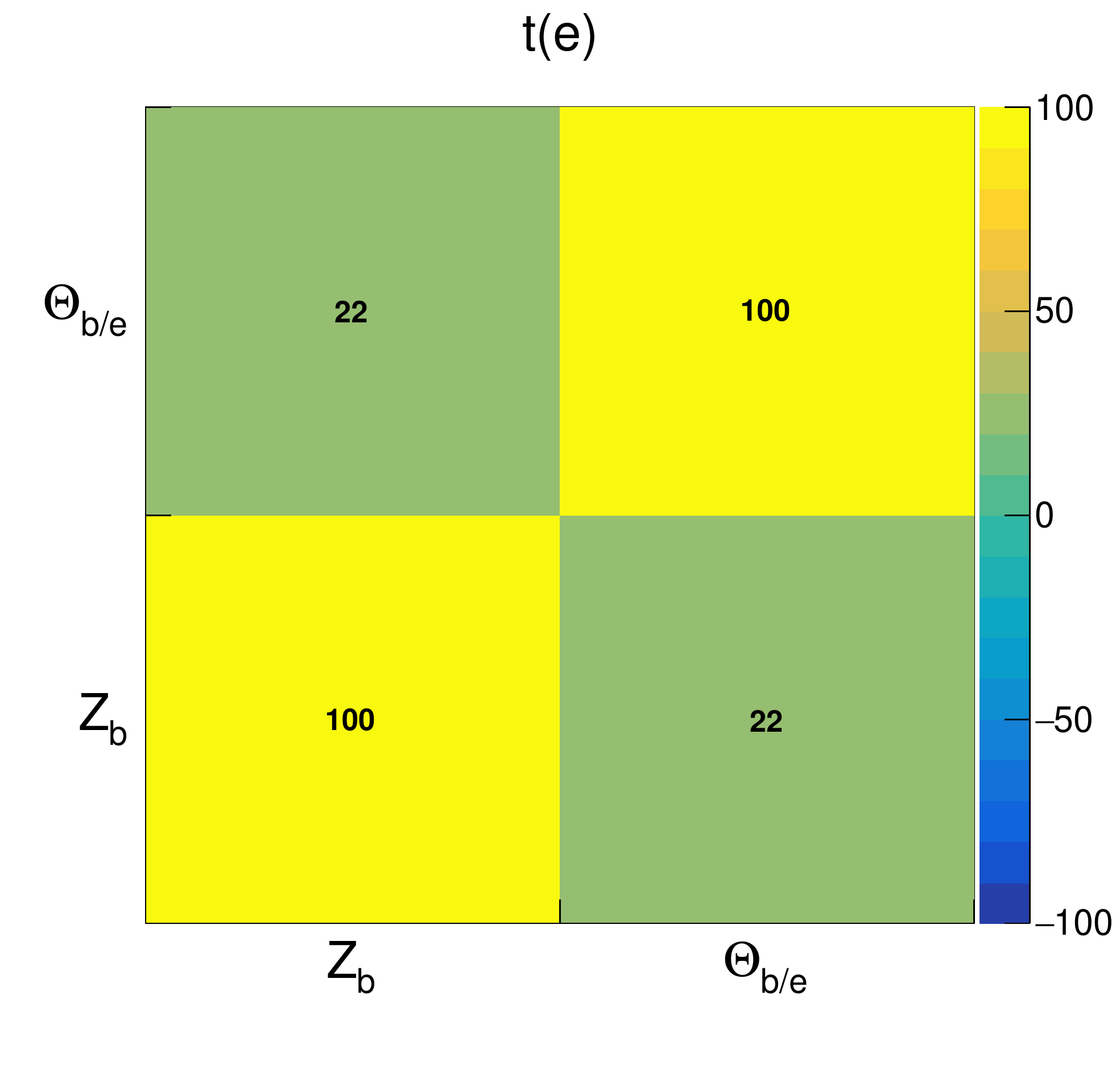}
		\caption{Linear correlation coefficients (in $\%$) between the variables in the sets $\mathcal{V}_e$ (left), and $\mathcal{V}_\nu$ (right) respectively, for the sample of top quarks decaying to electrons.}
		\label{fig:correlation}
	\end{center}
\end{figure}

In general, we expect a decent amount of correlations in studies with multiple variables. A useful way to represent it is via the linear correlation coefficient. To be specific, the  linear correlation coefficient   of  two variables $A$ and $B$, denoted by $\rho(A,B)$, is defined by the following equation 
\begin{equation}
\rho(A,B) \ \equiv \ \frac{E(AB) \ - \ E(A)E(B)}{\sigma(A) \sigma(B)}  \; , 
\label{Eq_Correlation}
\end{equation}
where $E(A)$, $E(B)$, and $E(AB)$  represents the expectation value of the variable $A$, $B$, and $AB$ respectively, and $\sigma(A)$ ($\sigma(B)$) stands for the standard deviation of $A$ ($B$). This quantity is useful to estimate the redundancy in information carried by the variables in a given set. For the variables listed in sets $\mathcal{V}_e$ and $\mathcal{V}_\nu$, we take all possible pairwise combinations (within each set) and calculate the  linear correlation coefficients for the sample of $t(e)$ jets. We show the matrices of correlation coefficients in Fig.~\ref{fig:correlation}, which depicts the efficacy in choosing the set of variables. 
 The observables are largely uncorrelated. There is only mild correlation present among the nonhadronic energy fraction $f_\text{1-h}$ and $\tau_{21}$.

\subsection{Multivariate analysis}

To quantify the discrimination power of our method as described in Sec.~\ref{Sec_BkgVeto}, we proceed to multivariate analysis using BDT with binary classification, as implemented in {\textsc Toolkit for Multivariate Analysis} \cite{Hocker:2007ht} within the {\tt ROOT} framework \cite{Antcheva:2009zz}. 
We weigh each of the samples so that they have exactly the same $p_T$ distribution for the leading jet in the event.
The parameters used in BDT are chosen as follows: 
{\tt NTress}, the number of trees in a forest, is taken as $1000$, the minimum percentage of training events required in a leaf node is taken as {\tt MinNodeSize}$=2.5\%$, the maximum depth of the decision tree is taken as {\tt MaxDepth}$=2$, we use {\tt Gradient Boost} algorithm~\cite{friedman2001} for boosting the decision tree with corresponding parameter {\tt Shrinkage}$=0.10$.

We put the variables in the sets $\mathcal{V}_e$ and $\mathcal{V}_\nu$  as inputs to two separate BDTs, mentioned as $\mathcal{B}_{e}^{t/b}$ and $\mathcal{B}_{\nu}^{t/b}$ respectively in Eq.~\eqref{Eq_BDT} for the classification training. We optimize both the BDTs using the sample of $t(e)$ jets as the signal and the sample of $b$ jets as the background.  We have  explicitly checked to make sure that none of the BDTs are overtrained.

\begin{figure}[h]
	\begin{center}
		\includegraphics[width=0.475\textwidth]{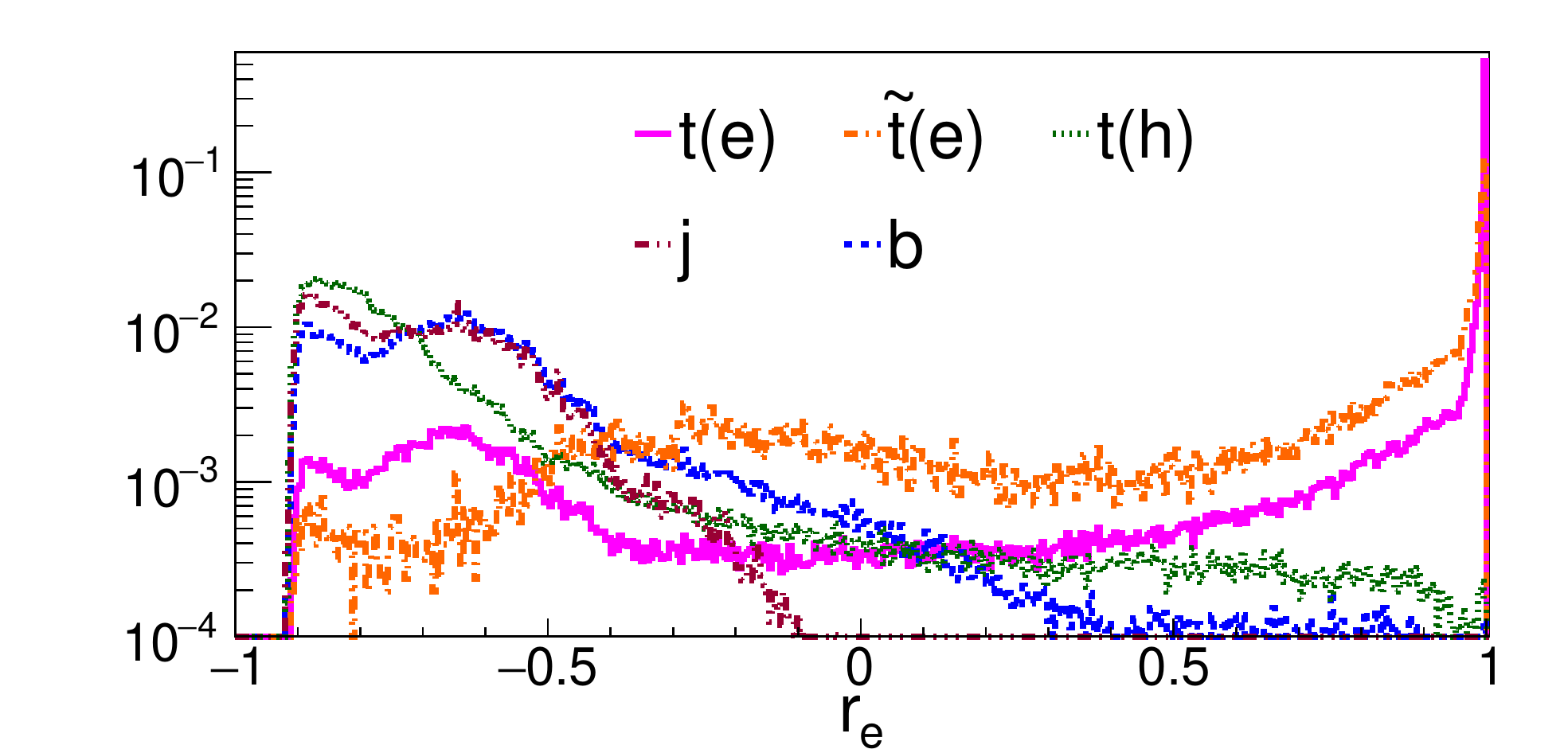}
		\includegraphics[width=0.475\textwidth]{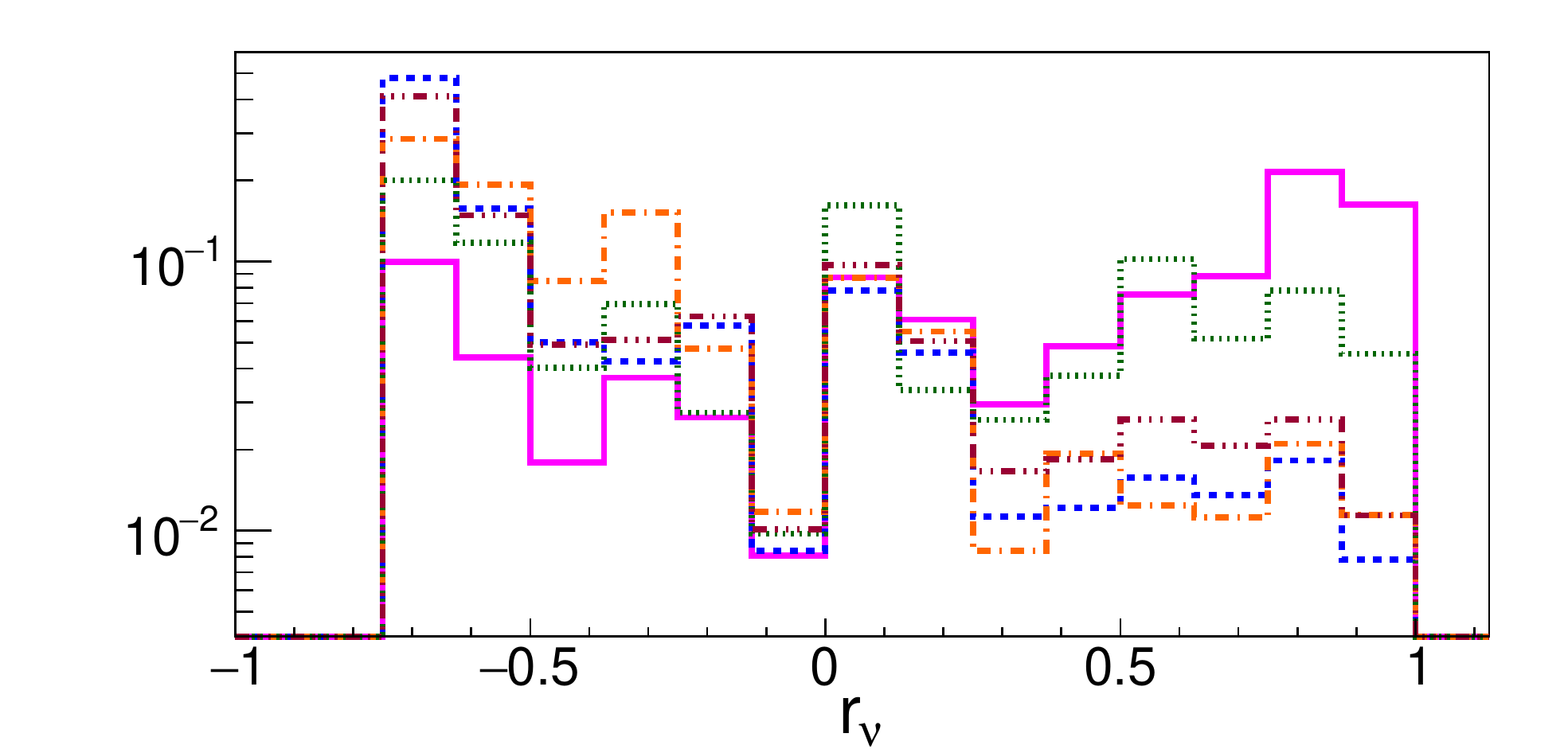}
		\caption{Distribution of BDT responses $r_e$ (left), and $r_{\nu}$ (right) respectively. See Eqs.~\eqref{Eq_BDT} and~\eqref{Eq_R} for the definitions of BDT responses.}
		\label{fig:BT1Ds}
	\end{center}
\end{figure}
\begin{figure}
	\begin{center}
		\includegraphics[width=0.475\textwidth]{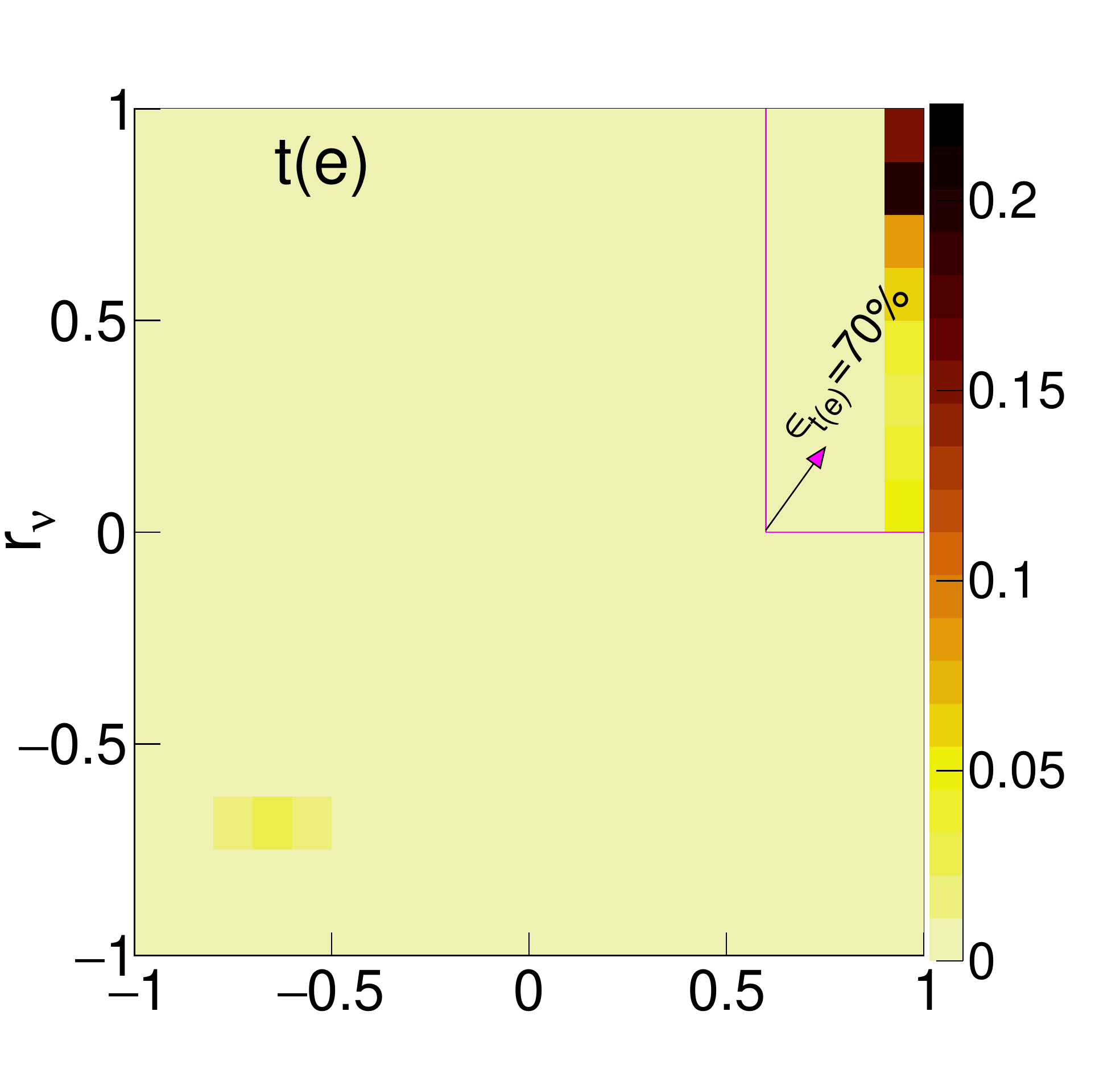}
		\includegraphics[width=0.475\textwidth]{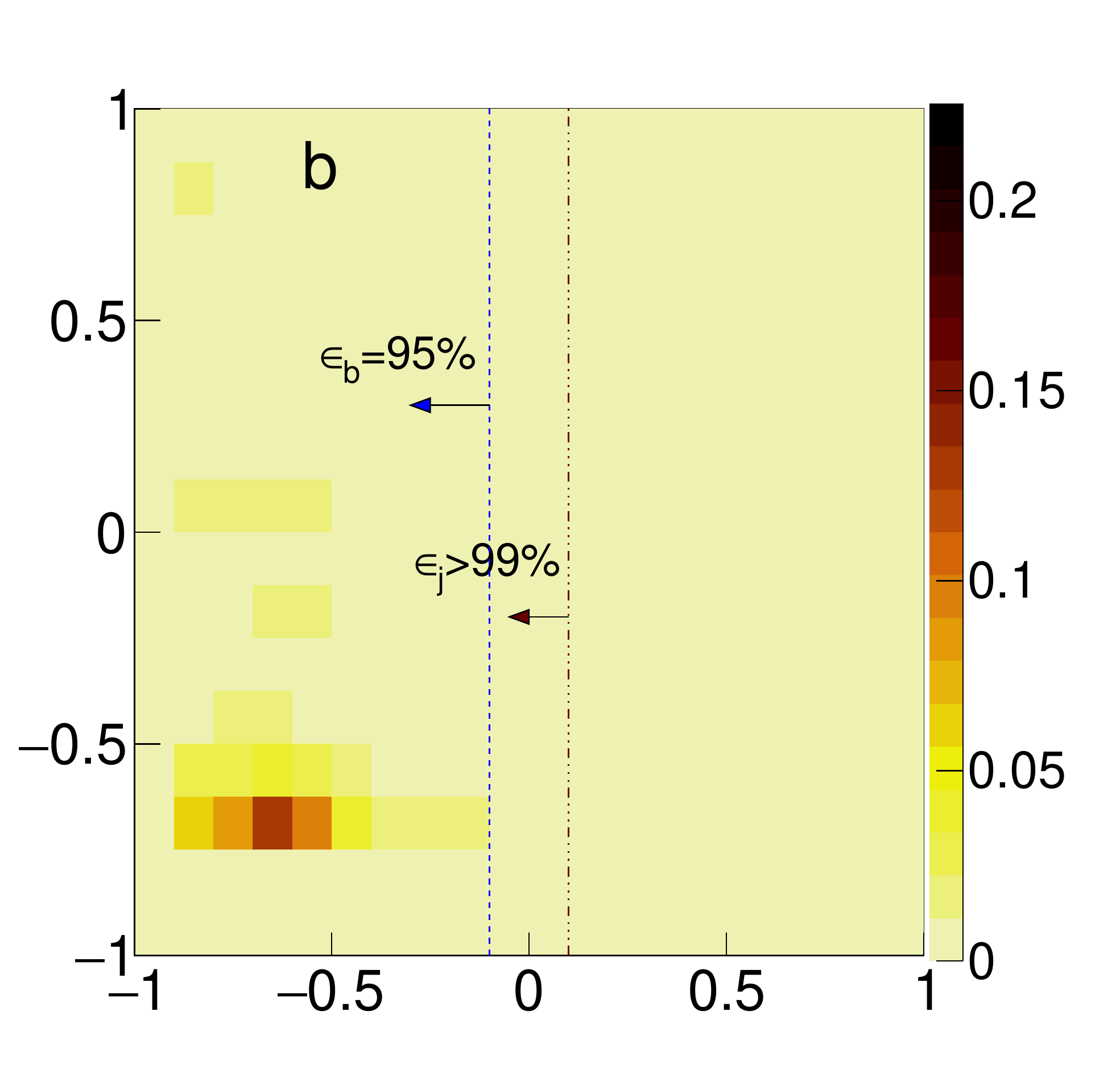}
		\includegraphics[width=0.475\textwidth]{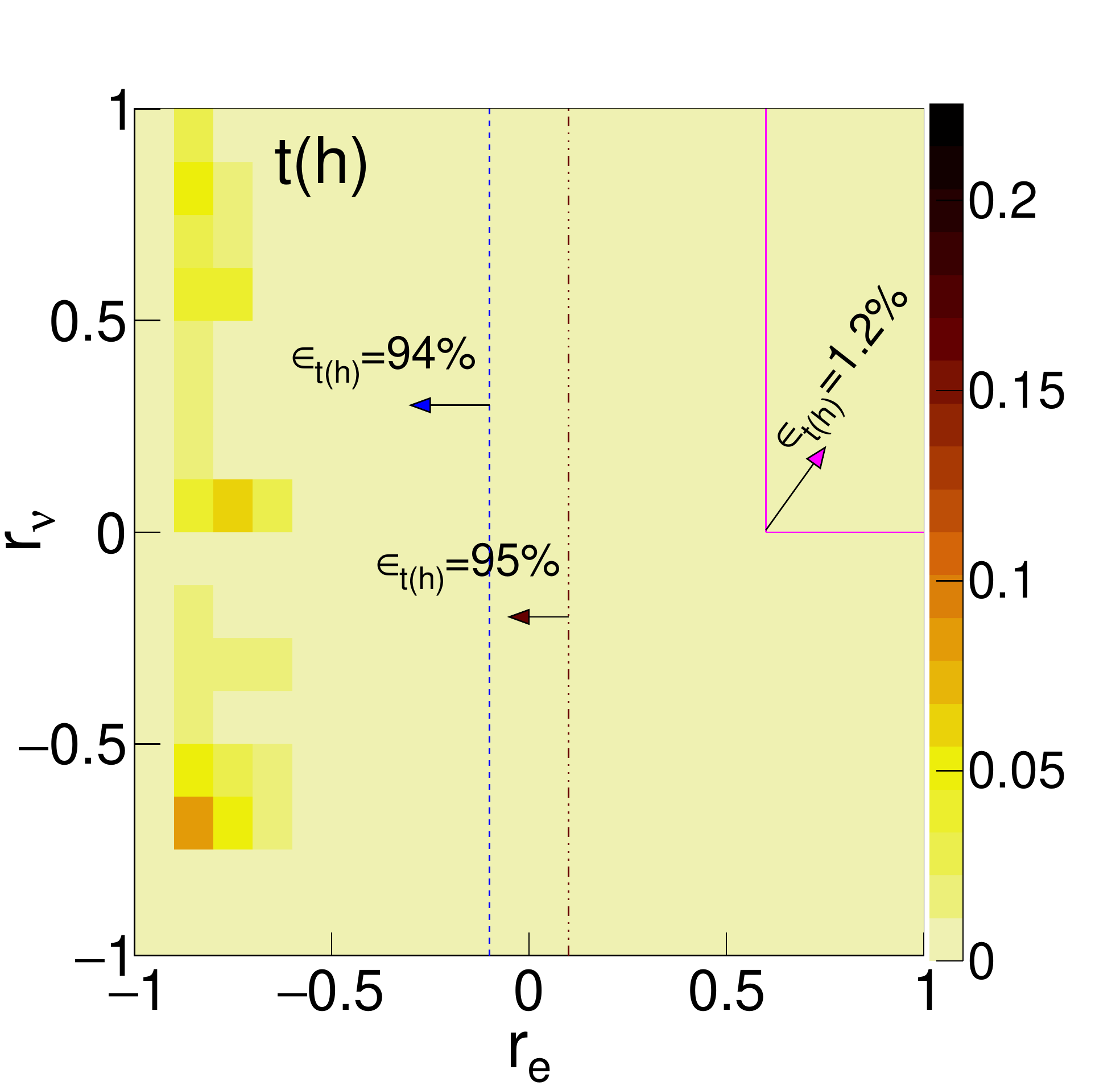}
		\includegraphics[width=0.475\textwidth]{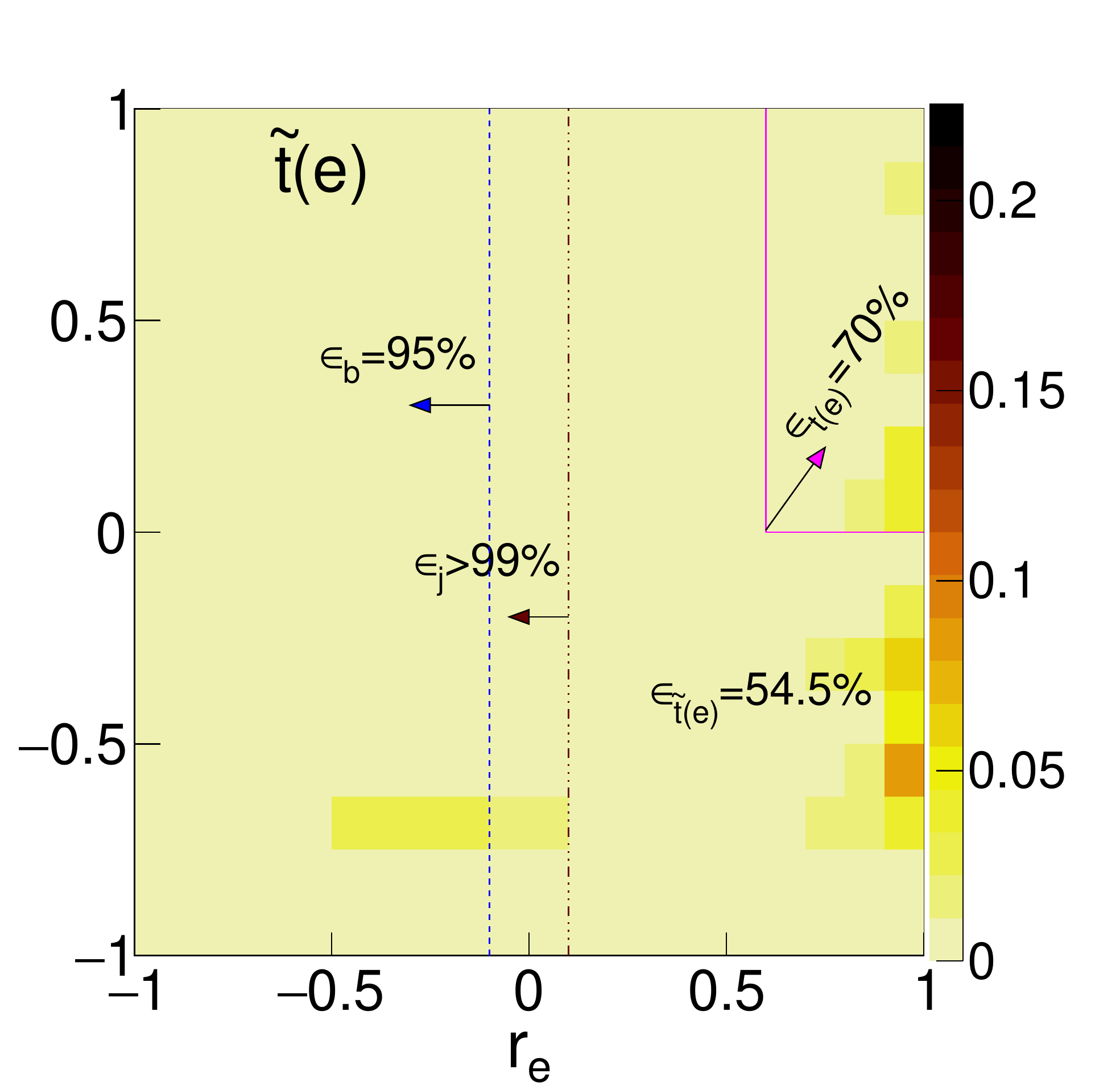}
		\caption{Probability distribution of BDT responses over a $2$-D plane for various event samples.
			Selection efficiencies for different samples are also overlaid in regions as defined in Sec.~\ref{sec:zone}.}
		\label{fig:BT2Ds}
	\end{center}
\end{figure}

Note, however, that once the BDTs are optimized, these simply can be treated as black boxes. The only purpose of these are then to map any jet to a number or a response.   See Eqs.~\eqref{Eq_BDT}--~\eqref{Eq_R} where we establish the notations for the BDTs ($\mathcal{B}_{e}^{t/b}$, $\mathcal{B}_{\nu}^{t/b}$) and their responses ($r_e,r_{\nu}$).  In Fig.~\ref{fig:BT1Ds} the probability distributions in $r_e$ and $r_\nu$ are shown for jets of different kinds. A far better understanding can be reached, however, by rather observing the joint probability distributions in the $r_e$--$r_\nu$ plane.  As explained before, once a jet is characterized by two responses (namely, $r_e$ and $r_\nu$), it is mapped to a point in the $2$-dimensional  plane of BDT responses $\{r_e, r_\nu \}$. We show the probability distributions of all types of jets in Fig.~\ref{fig:BT2Ds}, except for light flavor jets, since it is largely similar to $b$ jets (as expected from the distributions shown in Figs.~\ref{fig:variables_1}-\ref{fig:variables_3}, and Fig.~\ref{fig:BT1Ds}). 

As advertised before in Sec.~\ref{sec:method}, we find $t(e)$ jets around the corner $\left( +1, +1 \right)$  and  $b$ jets around the corner $\left( -1, -1 \right)$ and the separation is clear. As far as $t(h)$ jets are concerned, these are characterized by small $r_e$. This suggests that a simple use of $r_e$ may be sufficient to get rid of background jets due to $t(h)$. Note that this is a bonus feature since $\mathcal{B}_{e}^{t/b}$ is optimized to separate $t(e)$ jets from $b$ jets  and we did not use any additional information pertaining to the hadronic decay of top quarks. 
In the $r_{\nu}$ direction ($y$-axis), however, there is no clear separation of $t(e)$ jets from $t(h)$ jets. This is understandable, since as stated in Sec.~\ref{Sec_Results_Vnu}, we expect $t(h)$ jets to satisfy the critical ansatz made in Sec.~\ref{V_nu} often.    

Note that the true benefit of using variables in $\mathcal{V}_\nu$ and, consequently, of the response $r_{\nu}$ can be seen in probability distribution of $\tilde{t}(e)$ jets in Fig.~\ref{fig:BT1Ds}-\ref{fig:BT2Ds}. In this direction stop jets get completely separated from $t(e)$ jets. Not surprisingly, we see that $r_e$ fails to create reasonable separation between  $\tilde{t}(e)$ jets and $t(e)$ jets. 

Given the probability distributions in Fig.~\ref{fig:BT2Ds} we can construct a tagger for electronic top jets in two ways as described below.

\subsubsection{Cut on response $r_e$}

 We simply use cuts on $r_e$; this turns out to be a powerful discriminant for separating electronic top jets from background jets in the standard model, such as light flavored jets, QCD $b$ jets, and hadronic top jets. The main disadvantage of this procedure is that jets from new physics (in this case, jets containing remnants of boosted stop decays) may fake these at an alarming rate.   
	\begin{figure}[h]
		\begin{center}
			\includegraphics[width=0.475\textwidth]{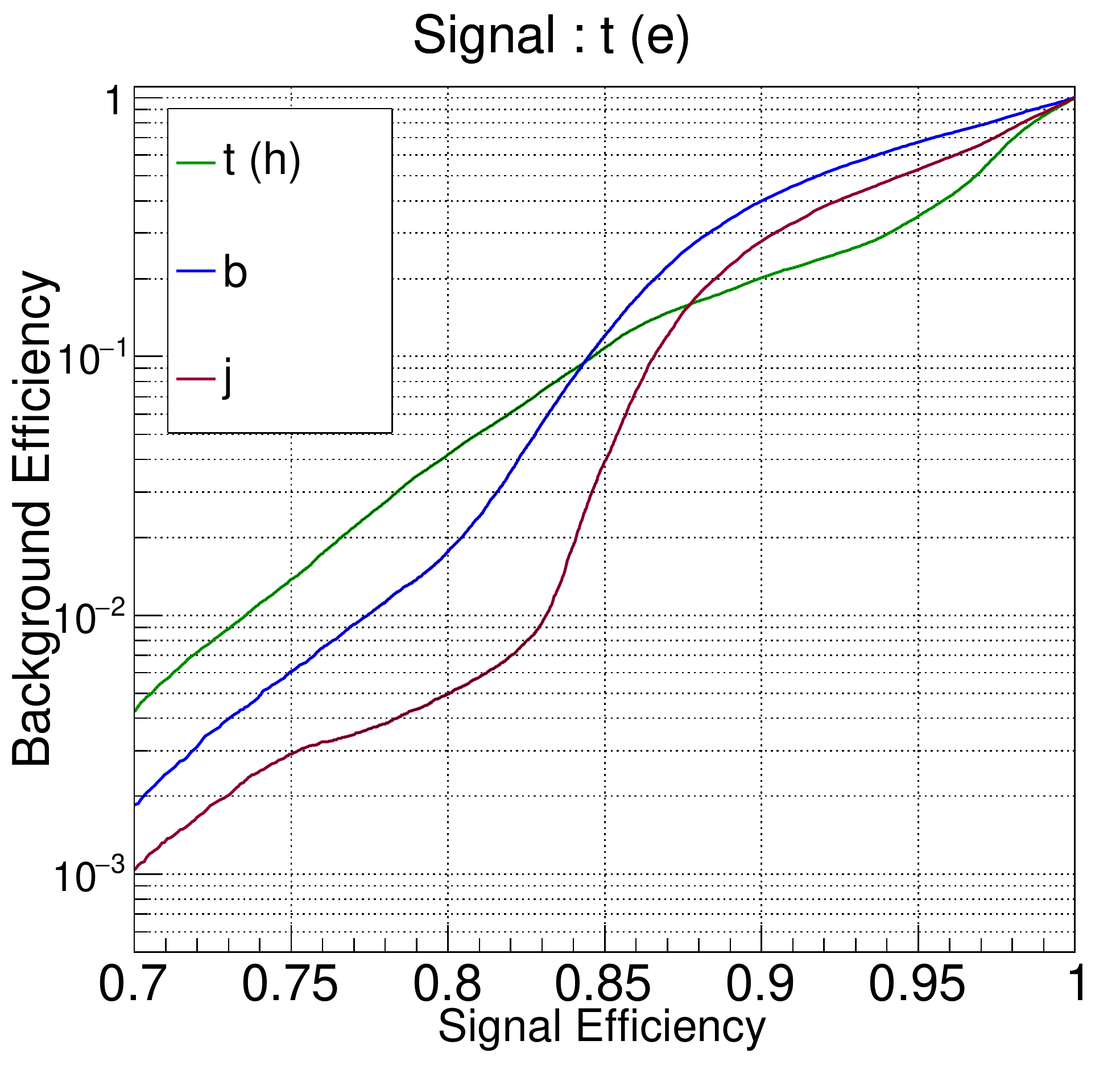}
			\caption{Receiver operator characteristic (ROC) curves quantifying the performance of the BDT classifier to identify top jets, where the top quark decays to a final state with electron, from various backgrounds.}
			\label{fig:ROCs}
		\end{center}
	\end{figure}
In order to benchmark the performance, we provide Fig.~\ref{fig:ROCs}, where we show the Receiver Operator Characteristic (ROC)  curves for signal ($t(e)$ jets) and various backgrounds. Note again, in order to produce this plot we only use $r_e$, the response of $\mathcal{B}_e$, the BDT trained to separate $t(e)$ jets from $b$ jets using only the observables in $\mathcal{V}_e$.  A cut on $r_e$, gives an acceptance for signal ($t(e)$ jets), as well as acceptances for other jets such as $b$ jets, light flavor jets, and $t(h)$ jets. Therefore, a single signal efficiency ($\epsilon_{t(e)}$) is associated with three different background efficiencies (namely, $\epsilon_b$, $\epsilon_{j}$ and $\epsilon_{t(h)}$ for the respective type of jets).  Consequently, we obtain three different ROCs at the same time (using only cuts on $r_e$).   As expected, we see for a top quark decaying to an electron, our tagger is able to provide fantastic separation: $\gtrsim 75\%$ signal efficiency at $1\%$ mistag rate from $b$ jets as well as light flavor jets and less than $2\%$ mistag rate from hadronic top jets.   

As long as we are only concerned with backgrounds from SM, clearly a single cut on $r_e$ is sufficient. However, as mentioned before, variables in $\mathcal{V}_{e}$ (and therefore $r_e$) only take into account the fact that the jet under consideration contains an energetic electron.  Any jet containing an energetic electron, whether in new physics events (from $\tilde{t}$ decays here) or even from SM events (because of kinematics) will most likely be misidentified as $t(e)$ jets by cuts on $r_e$ alone. This is why we do not show the mistag rates from   $\tilde{t}(e)$ jets in Fig.~\ref{fig:ROCs}.

\subsubsection{Construction of zones in response plane}
\label{sec:zone}
As argued before, the correct way to identify the electronic top jets, while at the same time to minimize the fake rate from other jets containing energetic electrons involves using more kinematic information associated with top quark decays (\textit{i.e.}, use $\mathcal{V}_\nu$). In our second approach, we consider the two dimensional probability distributions (in the $r_e$--$r_\nu$ plane) for signal and background jets as shown in Fig.~\ref{fig:BT2Ds}. It is clear that a rectangular cut that separates out the top-right corner in the $r_e$--$r_\nu$ plane yields a region dense in $t(e)$ jets. The large value of $r_e$ ensures small fake rate from $b$ jets, light flavor jets, and $t(h)$ jets, whereas a cut on $r_\nu$ simply gets rid of large  fraction of jets containing remnants of $\tilde{t}(e)$ decay.  We denote this region in the $r_e$--$r_\nu$ plane to be the electronic top or $t(e)$ zone. As shown in Fig.~\ref{fig:BT2Ds}, we suggest one such demarcation: 
\begin{equation}
	t(e)\text{ zone} \ \equiv \ r_{e}>0.6 \ \qquad \text{and} \qquad r_{\nu}>0  \; .
\end{equation}
We show the signal and background efficiencies for jets of different kinds in Table~\ref{tab:eff_2d}. At the operating point of $\gtrsim 70\%$ signal efficiency we find less than $1\%$ mistag rate from $b$ jets as well as light flavor jets and around $1.2\%$ mistag rate from hadronic top jets as before. However, note that we have established sufficient control even on jets due to stop quark decays (fake rate less than $20\%$). Of course, harder cuts on $r_\nu$ can yield even purer sample. 

An added benefit of this procedure is that one can demarcate a zone in the $r_e$--$r_\nu$ plane where one does not expect jets either due to light flavors or heavy flavors ($b$ jets, $t(e)$, as well as $t(h)$ jets). Jets that arise here can be termed as anomalous jets in the spirit of Ref.~\cite{Chakraborty:2017mbz}; in other words, these are less likely to be any of the standard jets.  The clue towards constructing the anomalous zone is also in Fig.~\ref{fig:BT2Ds}, which shows that $b$ jets and light flavored jets occupy mostly the left portions in the  $r_e$--$r_\nu$ plane. Therefore, using simple cut on $r_e$, as well as excluding the $t(e)$-zone one can find an anomalous zone: 
\begin{equation}
\text{Anomalous zone} \ \equiv \ \begin{cases}
\text{{\tt Case 1}} \ : \ \text{if } r_{\nu}<0, \quad  r_e > -0.1  \quad \text{else} \quad  -0.1 < r_e < 0.6 \\
\text{{\tt Case 2}} \ : \ \text{if } r_{\nu}<0, \quad  r_e > +0.1  \quad \text{else} \quad  +0.1 < r_e < 0.6 \\
\end{cases}
\end{equation}
The two cases we consider primarily differ in the lowest cut on $r_e$. As shown in Fig.~\ref{fig:BT2Ds}, the blue dotted line (the red dash-dotted line)  demarcates a zone that contains  $95\%$  of $b$ jets ($99\%$ light flavored jets). Consequently, we find that (shown in Table~\ref{tab:eff_2d}) {\tt Case 1} yields $3.8\%$ fake rate from $b$ jets and slightly more that $1\%$ fake rate from light flavor jets. {\tt Case 2} with more aggressive cuts gives much better numbers: around   $2.1\%$ fake rate from $b$ jets and less than $1\%$ fake rate from light flavor jets. 

\begin{table}[htbp]
\begin{center}
\begin{tabular}{ |c|c|c|c| }
\hline
\multirow{4}{*}{Efficiency}  & \multirow{4}{*}{$t(e)$ zone} & \multicolumn{2}{c|}{\multirow{2}{*}{Anomalous zone}}  \\
& &  \multicolumn{2}{c|}{ } \\
\cline{3-4}
&  & \multirow{2}{*}{ {\tt Case 1} } & \multirow{2}{*}{ {\tt Case 2}} \\ 
& & & \\
\hline \hline
$\epsilon_b$ & $<1\%$ & $3.8\%$ & $2.1\%$  \\
$\epsilon_j$ & $<1\%$ & $1.1\%$ & $<1\%$ 	\\
$\epsilon_{t(h)}$ & $1.2\%$ & $5.0\%$  & $3.4\%$ \\
$\epsilon_{t(e)}$ & $\mathbf{70\%}$ & $12.1\%$  & $10.4\%$	\\
$\epsilon_{\tilde{t}(e)}$ & $17.2\%$ & $\mathbf{60.0\%}$ & $\mathbf{54.5\%}$	\\
\hline
\end{tabular}
\end{center}
\caption{Efficiency values for all the jet samples for different cuts on the response plane.}
\label{tab:eff_2d}
\end{table}	

The two cases mentioned above warrants a thorough discussion in the philosophy of our proposed methodology, and especially the role $b$ tagging plays. Of course, to implement this proposal one requires a control sample of $b$ jets in order to optimize BDTs. After optimization, however, we may not need to impose $b$ tagging at all -- cuts on $r_e$ and $r_\nu$ are sufficient to reduce background. Take first the case of tagging electronic top jets. Clearly, the fake rate in the $t(e)$ zone is less than $1\%$ even for light flavor jets without requiring any further $b$ tagging. The anomalous zone in {\tt Case 2} also provides another example where even without $b$ tagging the  mistag rate from light flavored jets can be controlled well below $1\%$.  Even in    {\tt Case 1}, the anomalous zone has $\epsilon_j = 1.1\%$, which suggests that this operating point can be useful as long as the fake rate from $b$ jets at order $4\%$ is tolerable. 

Before concluding, let us note that, as the name suggests, the construction of ``anomalous zone" is not the same as tagging stop particles decaying to electrons. It rather finds anomalous objects which are less likely to be either light flavor jets or heavy flavor jets (including top jets). Once a set of events are identified to contain these jets, one can look into the global event information and attempt to unearth the underlying physics giving rise to these objects. In this work, we take jets containing decay products of stop particle as an example of such anomaly and demonstrate that it can be found at good efficiency. We, however, emphasize that the anomalous zone as constructed here is not influenced by the properties of the new physics particle. We simply identify a region, rather rare in standard jets -- it is  model-independent.  The rate at which new physics jets can be captured in this zone, of course, is ultimately going to vary depending on the very nature of the new physics processes. 
It will be also interesting to explore the dependence of the efficacy of the methods presented here, on the polarization of the top quark, which as discussed earlier, can be a great probe of BSM physics in the top quark production.

\section{Conclusion}\label{sec:conclusion}

In this paper, we presented a new method to identify jets consisting of all the visible remnants of a boosted top quark which decays to an electron.  The first part of this method uses observables computed for a large sized jet using  information from different parts of the detector such as the  tracker,  the electromagnetic calorimeter and the hadron calorimeter, with which one tries to determine whether the jet is consistent with a jet containing an energetic electron. 
In the second part of this methodology, we propose a way to identify the four-momentum associated with the candidate for electron inside the jet. Since  the electron shower overlaps with the shower initiated by the $b$ quark  within these jets, identification of the electron becomes hard. Even if an electron inside a jet is identified, it is difficult to pinpoint whether the electron rich jet is indeed due to top quark decay or not, because the invisible neutrino carries away a nontrivial part of the energy-momentum of the original top quark.
To this end we suggest  one more step in the construction of a tagger for a top quark decaying to a final state with an electron.
 We employ an extra  ansatz that there exists a massless invisible  four-momentum, roughly collimated with the electron, which when combined with the electron and the full jet, reconstructs the four-momentum of an on-shell $W$ boson and top quark respectively. This allows us to determine some features of the top quark decay kinematics given entirely in terms of visible objects. Combining both these parts we construct a tagger which identifies electronic top jets at high efficiency, with small (controllable) mistag rate from background jets such as light flavor jets, $b$ jets, hadronic top jets, and even jets containing electrons inside (but not due to top quark decay). 


\begin{acknowledgments}
We thank the Workshop on High Energy Physics Phenomenology (WHEPP) 2017 where this work kick-started.  We also acknowledge the second Workshop of the  Indo-French Network  in High Energy Physics (LIA THEP and CEFIPRA INFRE-HEPNET) held in IISER Pune for creating the scope for lively and fruitful discussion.  We are thankful to the grid computing facility at DHEP, TIFR which has been used for Monte Carlo simulations.  We thank Nishita Desai for careful reading of this paper and providing useful suggestions.  This work was supported in part by the CNRS LIA-THEP and the INFRE-HEPNET of CEFIPRA/IFCPAR (Indo-French Centre for the Promotion of Advanced Research).  The work of RMG is supported by the Department of Science and Technology, India under Grant No. SR/S2/JCB-64/2007.  TSR was supported in part  by  the Early Career Research Award by Science and Engineering Research Board, Dept. of Science and Technology, Govt. of India (grant no. ECR/2015/000196).  SC thanks Sabyasachi Chakraborty for technical help during the very initial phase of the project, and Aravind H. V., Soham Bhattacharya, and Gobinda Majumder for useful discussions.
\end{acknowledgments}


\appendix

\section{Appendix 1}\label{sec:appendix1}
\begin{figure}[hbtp]
	\begin{center}
		\includegraphics[width=0.243\textwidth]{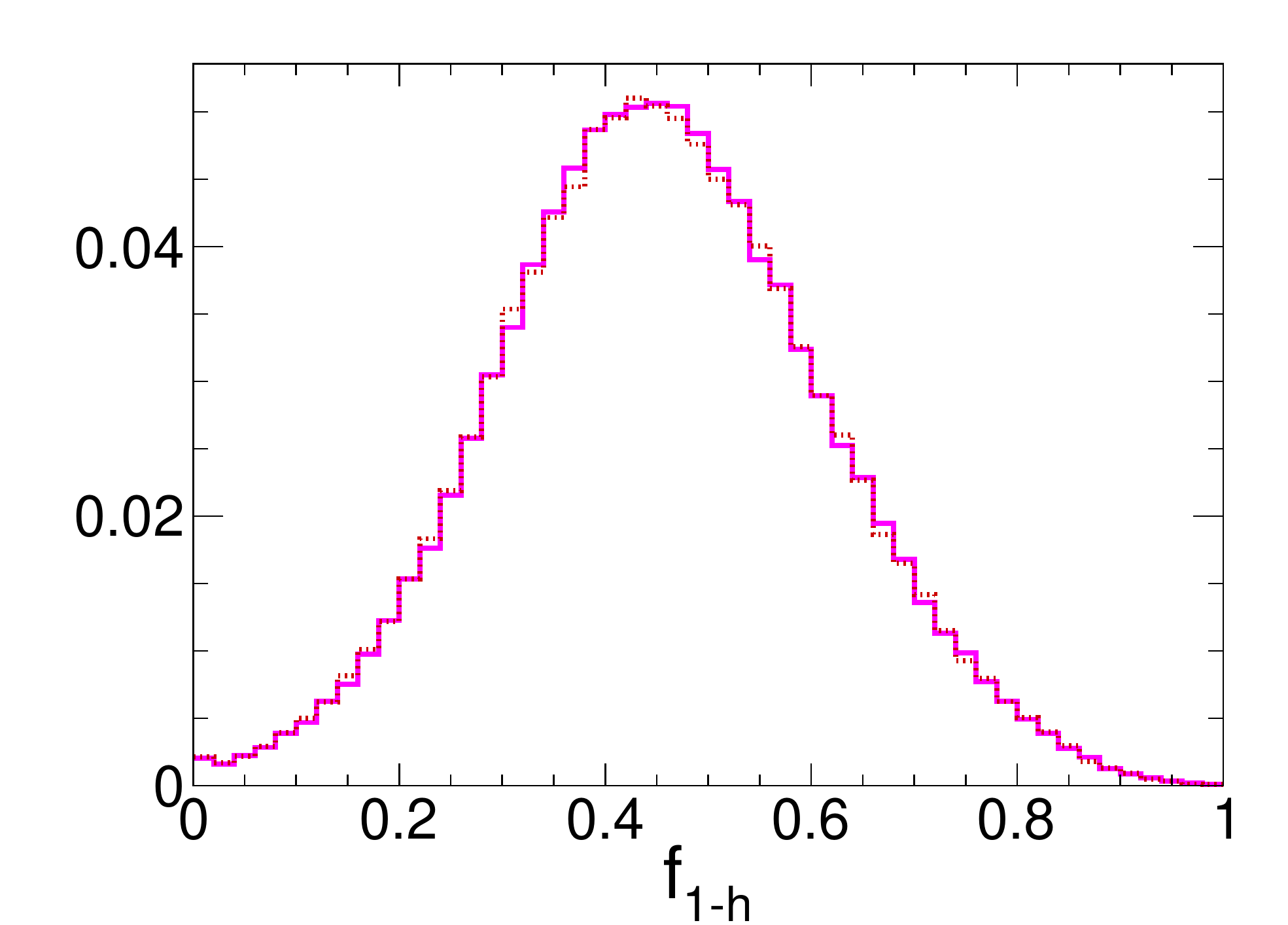}
		\includegraphics[width=0.243\textwidth]{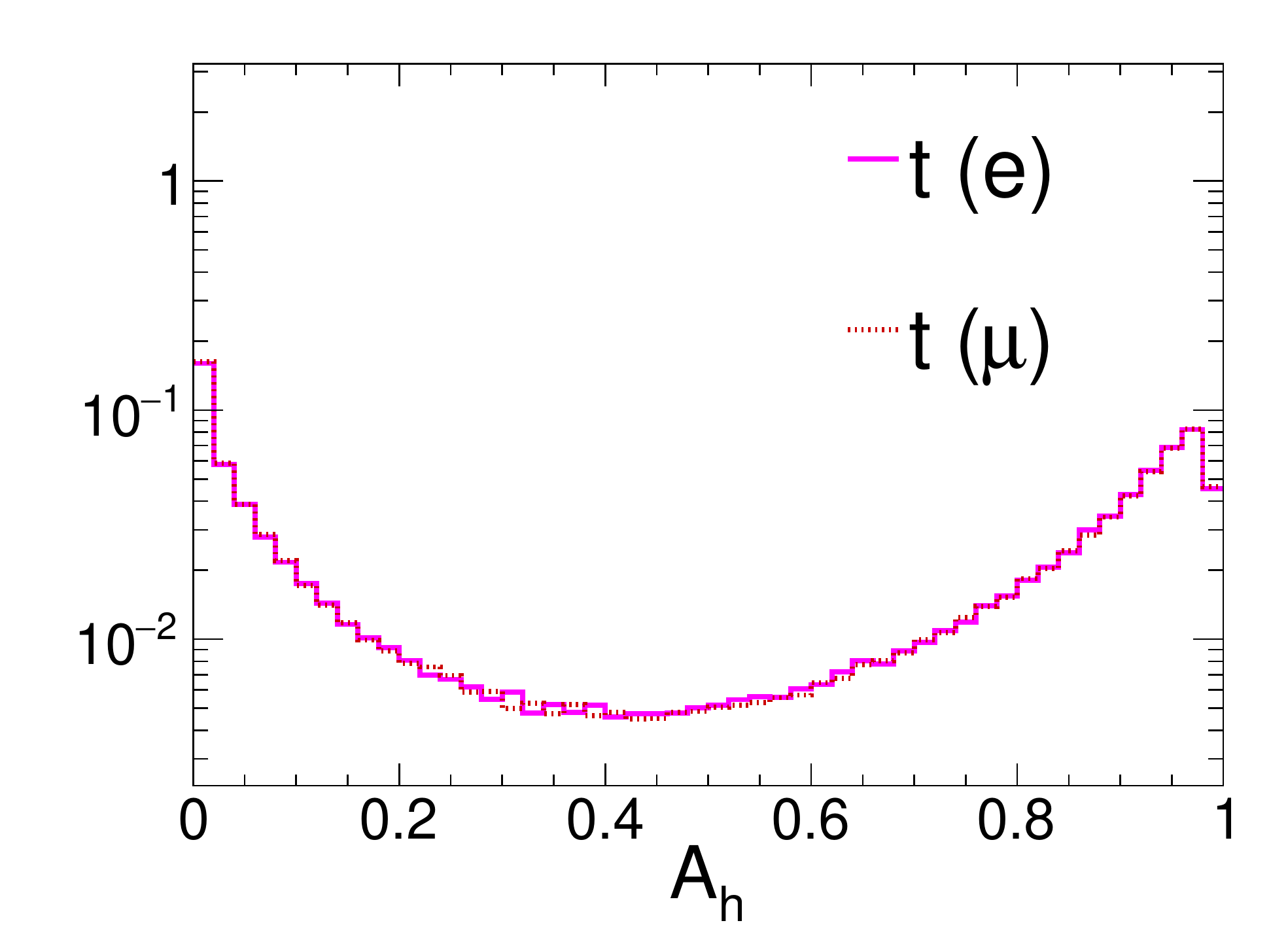}
		\includegraphics[width=0.243\textwidth]{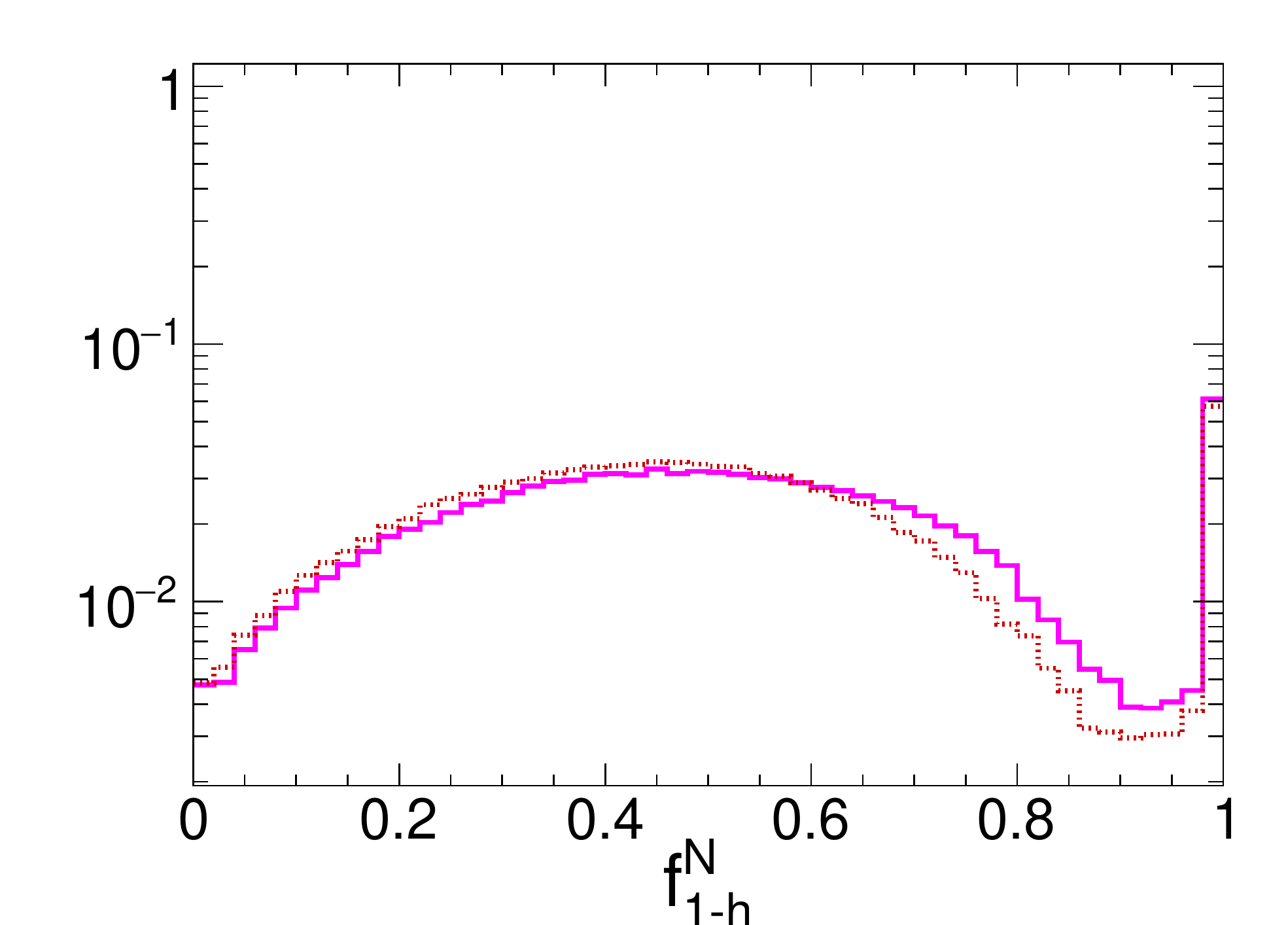}
		\includegraphics[width=0.243\textwidth]{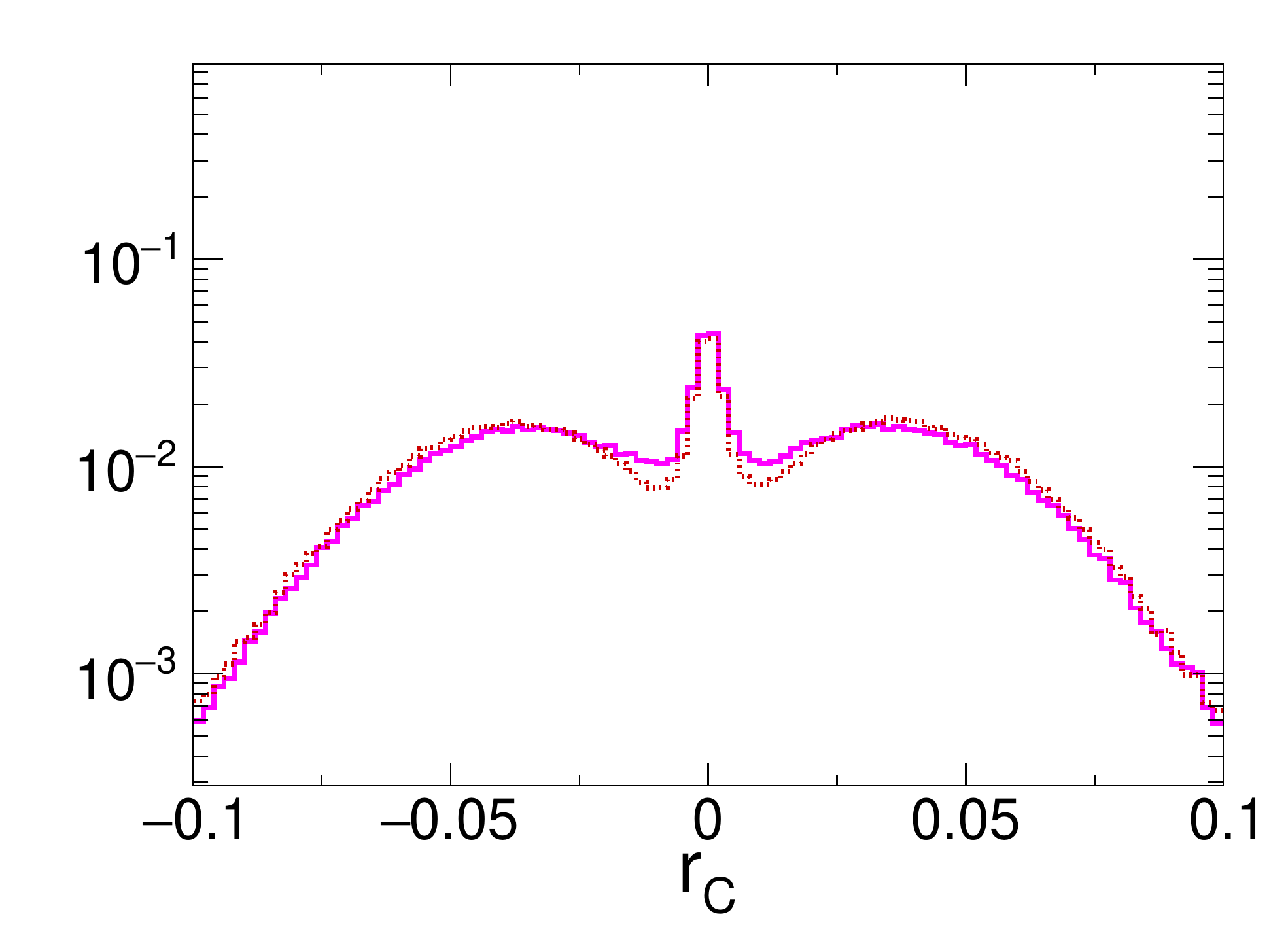}
		\includegraphics[width=0.243\textwidth]{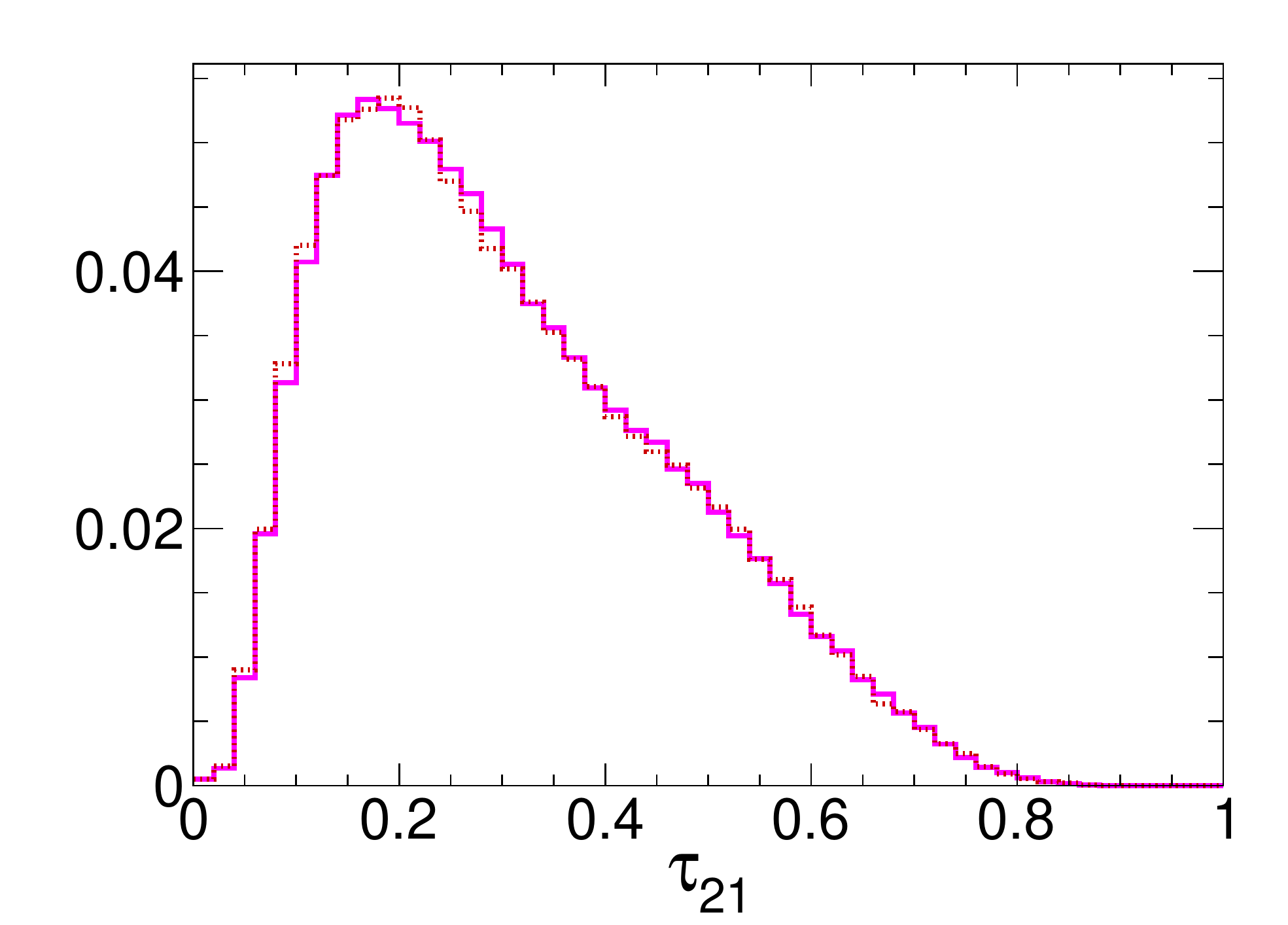}
		\includegraphics[width=0.243\textwidth]{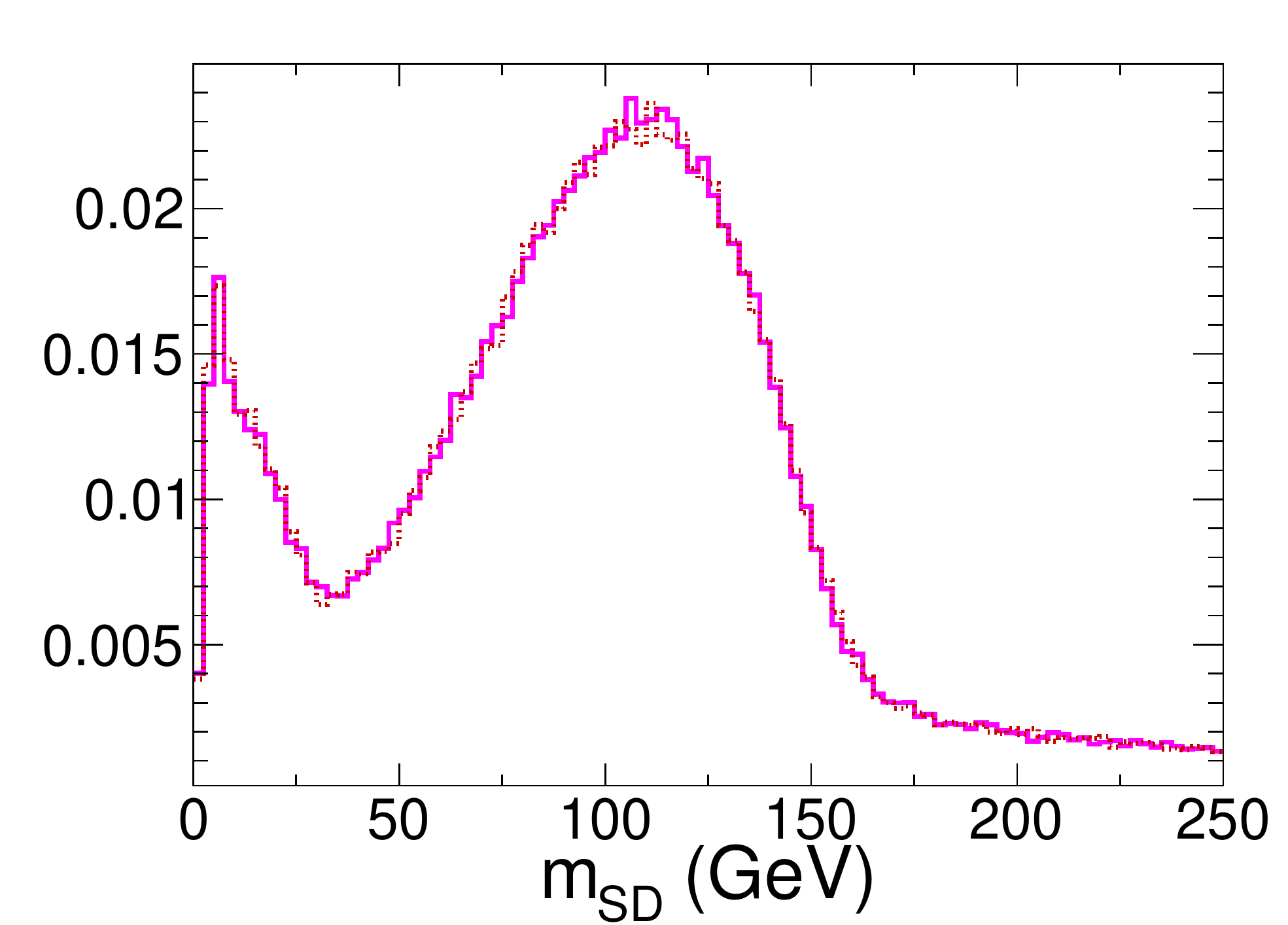}
		\includegraphics[width=0.243\textwidth]{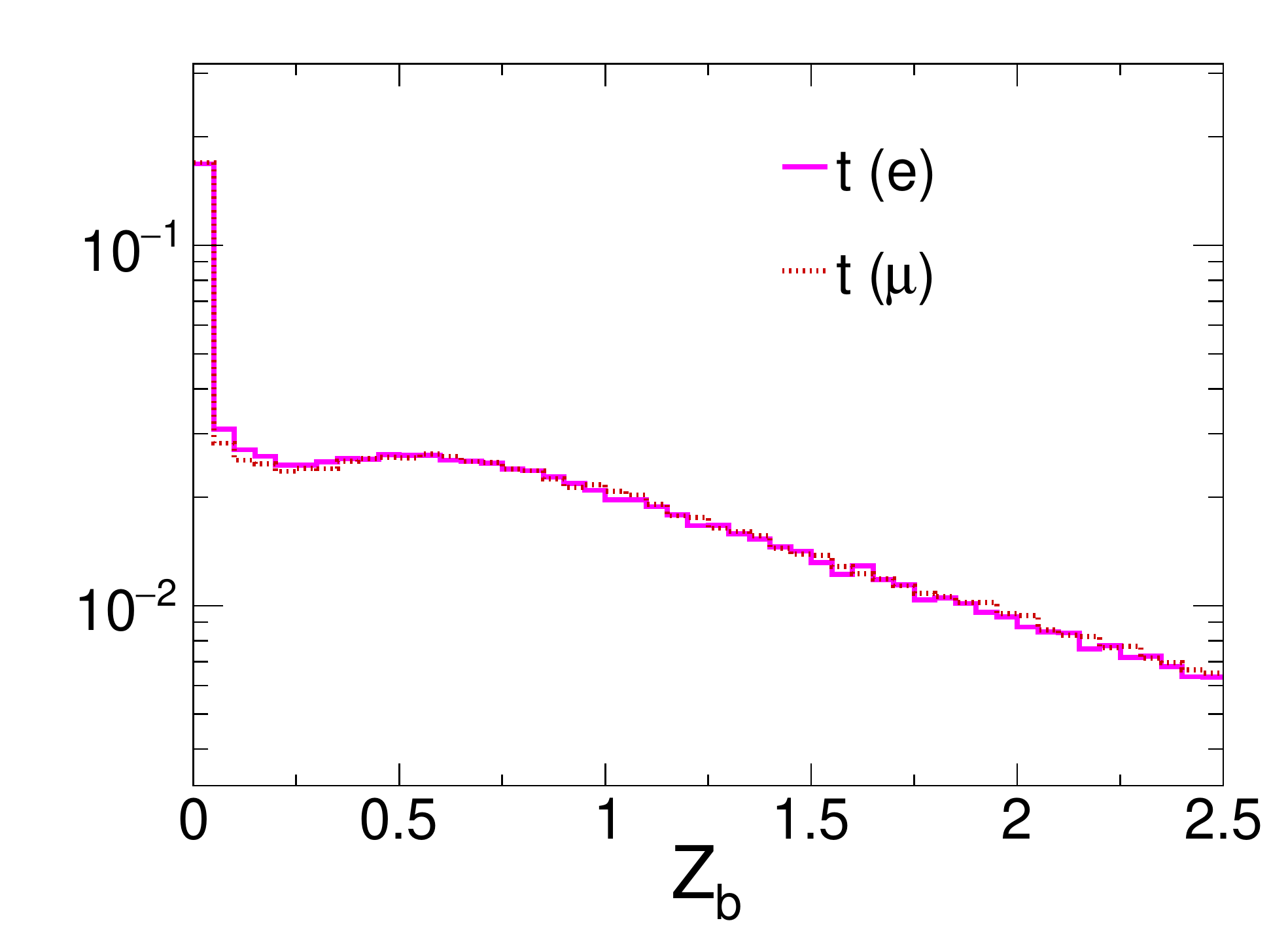}
		\includegraphics[width=0.243\textwidth]{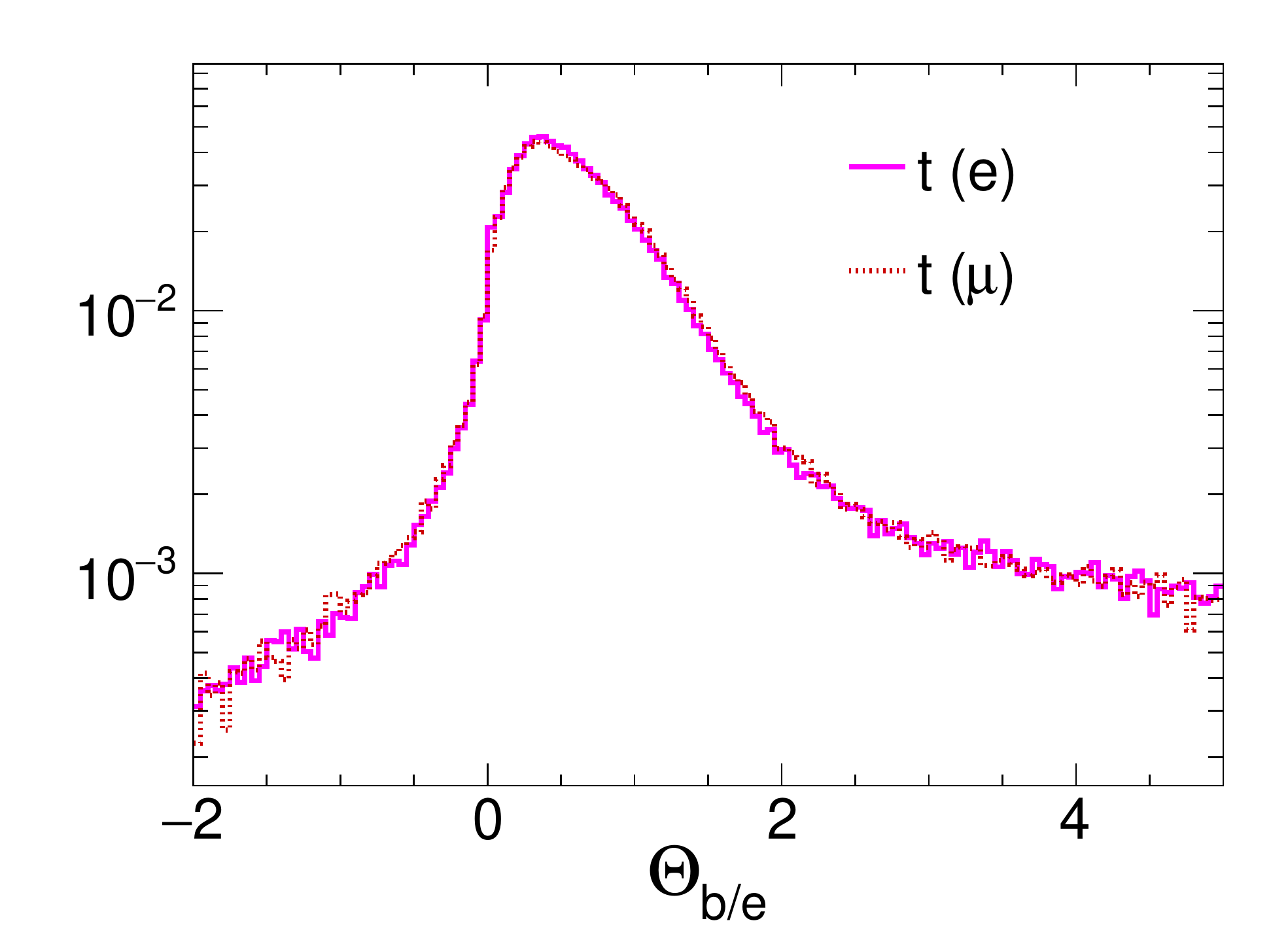}
		\caption{Distribution of $f_\text{1-h}$, $A_h$,  $f_\text{1-h}^N$, $r_C$, $\tau_{21}$, $m_{SD}$, $Z_{b}$, and $\Theta_{b/e}$ for $t(e)$ and $t(\mu)$ jets.}
		\label{fig:Variables_temu}
	\end{center}
\end{figure}
We demonstrate in this appendix that the methodology we propose in constructing electronic top tagger works equally well when one considers the top quark decays to a muon. Note that as described in Section~\ref{sec:sample}, jets are constructed using particle flow elements from Delphes and therefore contain muons. All the variables included  in the set $\mathcal{V}_{e}$, should remain the same, by construction, whether the top quark decays to an electron or a muon. The only difference stems from how we find the muon candidate. As argued before, particle flow muons are significantly more reliable after one matches the track observed in the tracker  with that at the muon spectrometer.  In this work we simply take the four-momentum associated with the most energetic particle-flow muon among the jet constituents as the four-momentum of the muon candidate, with which we calculate both the observables in the $\mathcal{V}_{\nu}$ set.   
We show the comparison of all the variables in $\mathcal{V}_e$ and $\mathcal{V}_{\nu}$ for electronic top jets and the muonic top jets in Fig.~\ref{fig:Variables_temu}. Apart from some minor deviations (associated with more radiation in case of electrons), \emph{all} the distributions in  electronic vs. muonic top jets are identical. We also explicitly checked and found that the BDT responses are very similar for these two jet samples. Therefore, the methodology proposed here can find muonic decays of top at least as efficiently as electronic top jets using the same cuts proposed here.

\bibliographystyle{jhep}
\bibliography{leptop}

\end{document}